\newcommand{\ergs}{\ensuremath{\rm\,erg\,s^{-1}}}
\newcommand{\phigh}{\ensuremath{P_{1.4}}}
\newcommand{\shigh}{\ensuremath{\sigma_{1.4}}}

\newcommand{\meanpcav}{\ensuremath{\langle P_{\rm{jet}} \rangle}}
\newcommand{\meanpcavzlx}{\ensuremath{\langle P_{\rm jet}(z_i,L_{\rm X})\rangle}}
\newcommand{\pcav}{\ensuremath{P_{\rm{jet}}}}
\newcommand{\tcool}{\ensuremath{t_{\mathrm{cool}}}}
\newcommand{\kpc}{\mbox{\ensuremath{\mathrm{~kpc}}}}

\documentclass{emulateapj}

\usepackage{times}

\shorttitle{Mean jet power in X-ray clusters}
\shortauthors{Ma et al.}

\begin{document}

\defcitealias{ma11b}{MMN11}

\title{Radio AGN in galaxy clusters: heating hot atmospheres and
  driving supermassive black hole growth over cosmic time}

\author{
C.-J. Ma\altaffilmark{1,2},
B. R. McNamara\altaffilmark{1,2,3},
P. E. J. Nulsen\altaffilmark{2},
}
\altaffiltext{1}{Department of Physics \& Astronomy, University of
  Waterloo, 200 University Ave. W., Waterloo, Ontario, N2L 3G1,
  Canada.}
\altaffiltext{2}{Harvard-Smithsonian Center for Astrophysics, 60
  Garden St., Cambridge, MA, 02138-1516, United States.}
\altaffiltext{3}{Perimeter Institute for Theoretical Physics, 31
  Caroline St. N., Waterloo, Ontario, N2L 2Y5, Canada.}
\begin{abstract}

We estimate the average radio-AGN (mechanical) power deposited into
the hot atmospheres of galaxy clusters over more than three quarters
of the age of the Universe.  Our sample was drawn from eight major
X-ray cluster surveys, and includes 685 clusters in the redshift range
$0.1 < z < 0.6$ that overlap the area covered by the NVSS.  The
radio-AGN mechanical power was estimated from the radio luminosity of
central NVSS sources, using the relation of
\citet{cavagnolo10} that is based on mechanical powers
  determined from the enthalpies of X-ray cavities.  We find only a weak correlation between radio luminosity and cluster X-ray luminosity, although the most powerful radio sources resides in luminous clusters.
The average AGN mechanical power of $3\times10^{44}$\ergs\  exceeds the X-ray luminosity of $44\%$ of the clusters, indicating that the accumulation of radio-AGN energy is significant in these clusters. Integrating the AGN mechanical power to redshift $z=2.0$, using simple models for its evolution and disregarding the hierarchical growth of clusters, we find that the AGN energy accumulated per particle in low luminosity X-ray clusters exceeds $1$\,keV per particle. 
This result represents a conservative lower limit to the accumulated thermal energy. The estimate is comparable to the level of energy needed to ``preheat'' clusters, indicating that continual outbursts from radio-AGN are a significant source of gas energy in hot atmospheres.  Assuming an average mass conversion efficiency of $\eta=0.1$, our result implies that the supermassive black holes that released this energy did so by accreting an average of $\sim 10^9 M_\odot$ over time, which is comparable to the level of growth expected during the quasar era.

\end{abstract}


\section{Introduction}\label{sec:intro}

Models for the formation and evolution of cosmic structure generally
invoke some heating mechanism to prevent catastrophic gas cooling and
excessive star formation in massive galaxies
\citep[e.g.,][]{sijacki06}.  In galaxy clusters, the same mechanism
may generate the excess entropy responsible for deviations from the
self-similar scaling relations expected otherwise
\citep[e.g.,][]{markevitch98}.  For example, the $L_{\rm X}$ -- $T$
relation for galaxy groups is steeper \citep[$L_{\rm X} \propto
  T^{2.6}$;][]{markevitch98} than for clusters \citep[$L_{\rm X}
  \propto T^{2}$;][]{kaiser86,arnaud99}.  Excess entropy is also
revealed by flatter core entropy profiles in galaxy groups than in
massive clusters \citep[e.g.,][]{voit05}.  Furthermore, in cooling
core clusters, the heating rate needs to be related to the high rate
of radiative cooling that was previously thought to cause cooling
flows \citep{fabian94}.  The heating cannot be too effective, since
cooling cores are found in a large fraction of local X-ray clusters
\citep[e.g.,][]{mittal09, hudson10, santos10}, but it must be
sufficient to explain the scarcity of cooling gas that would be
expected to accompany strong cooling flows
\citep[e.g.,][]{peterson03}.

One of the most promising sources of this non-gravitational energy is
active galaxy nuclei \citep[AGN; e.g.,][]{mcnamara00}. The questions
remain of when, how and and how much AGN energy is distributed into
their environments \citep[e.g.,][]{short12,young11}.  The preheating
model \citep{evrard91,kaiser91} proposes that energy injected into the
intergalactic medium at high redshifts explains the observed
departures from self-similar scaling relations.  \citet{wu00} found
that the minimum excess energy required to break self-similarity is
$\simeq 1$\,keV/particle.  At high redshifts, many AGN are in the
radiatively efficient ``quasar'' mode \citep{croton06}, when high AGN
accretion rates are promoted by the high galaxy merger rate.  Although
most of the energy output of these AGN is radiated away, quasars are
so powerful that only a small fraction of this energy is required to
produce the excess entropy in hot atmospheres.  By contrast, there are
far fewer quasars at lower redshifts, but X-ray observations have
revealed that AGN in ``radio mode'' deposit significant amounts of
energy into their hot atmospheres.  The total energy output of AGN in
radio mode is generally less than in quasar mode.  Nevertheless, a
large proportion of this emerges as mechanical energy in jets and
simulations suggest that the radio mode feedback is necessary, in
addition to the preheating, to suppress cooling flows in clusters and
star formations in galaxies \citep[e.g.,][]{bower06, croton06,
  sijacki07}.  Radio AGN hosted by cluster central galaxies in the
local Universe have been shown to deposit enough power to prevent
rapid cooling and star formation in the centers of many clusters
\citep[e.g.,][]{birzan04, best07a}. Supported by the correlation
between radiative cooling rates in clusters and the radio power of the
central AGN \citep[e.g.,][]{rafferty06,dunn06}, the power output of
the AGN is believed to be coupled to the cooling rate of the hot gas
in a feedback loop \citep[][]{mcnamara07}.  Nevertheless, some
powerful AGN apparently reside in non-cooling core clusters
\citep[e.g.,][]{sun07}.  Although these systems lack large scale
cooling flows, accretion from small coronae around their AGN can
support powerful radio sources \citep[e.g.,][]{hardcastle07}.  Radio
AGN in the non-cooling core clusters have been shown to contribute
significantly to the excess entropy of less massive clusters
\citep[e.g.,][\citetalias{ma11b} hereafter]{best07a,giodini10, ma11b}. 

In this paper, we focus on estimating the average mechanical power
output of radio AGN in clusters out to $z \simeq 0.6$, corresponding
to a look-back time of about 5.7\,Gyr.  At these redshifts, it is
difficult to estimate directly from X-ray images the amount of energy
deposited by AGN in the intracluster medium (ICM).  The only
systematic search for X-ray cavities at these redshifts was undertaken
recently by \citet{hlavacek12}.  They concentrated on identifying
cavities in the most luminous, and bright cooling core clusters and so
could not quantify the average AGN output of clusters overall, which requires a survey of clusters with and without
  cavities.
The question of how much energy AGN contribute to the ICM at higher
redshifts is significant because AGN activity increases with redshift
\citep[e.g.,][]{martini09}.  \citet{hart11} suggest that the power
injected into clusters by radio AGN at redshift 1.2 is substantial, a
factor of 10 greater than injected locally.  In addition, the fraction
of cooling core clusters appears to evolve with redshift, so that many
fewer large cooling cores are found beyond $z\sim0.3$ (e.g.,
\citealt{santos10, samuele11}; but see \citealt{santos12}).  If jet
power is coupled to cooling power by a feedback loop, this suggests
that the mean jet power in high-redshift clusters should be reduced.

In \citetalias{ma11b}, we estimated the
average mechanical energy deposited by radio AGN in galaxy clusters
using the clusters in the 400 Square Degree Cluster Survey
\citep[400SD, $0.1<z<0.6$;][]{burenin07} and the radio sources in the
NRAO VLA Sky Survey \citep[NVSS;][]{nvss} to show that the AGN feedback
in radio mode could also contribute significantly to the energy budget
of clusters and groups.  We found that 30\% of the clusters showed
radio emission within a projected radius of 250\,kpc and above a flux
threshold of 3 mJy, despite the declining numbers of cooling core
clusters in the 400SD \citep[][see also \citealt{mcnamara12};
  \citealt{mann12}]{santos10,samuele11}. The average jet power of the
central radio AGN is approximately $2\times10^{44} \ergs$.  Assuming
that the current AGN input power remains constant to redshifts of 2,
the energy input per particle would be at least 0.4\,keV within
$R_{500}$.  In addition, we found no significant correlation between
the radio power, i.e., the mechanical jet power, and the X-ray
luminosities of clusters in the redshift range 0.1 -- 0.6.  This
implies that the mechanical jet power per particle is higher in
clusters with lower masses.  However, within this single flux-limited
cluster survey, the X-ray luminous clusters are also the clusters with
the highest redshifts.  Thus, we could not distinguish redshift
evolution from luminosity dependence for AGN feedback.  In the present
study, we try to break this degeneracy using a composite sample from
eight X-ray cluster surveys.

The method used to estimate jet mechanical powers from the radio
powers of cluster central galaxies is reviewed in
\S\ref{sec:equation}.  The composite cluster sample is introduced in
\S\ref{sec:sample}.  In \S\ref{sec:fluxcorr}, we examine the
correlation between the power of a radio galaxy and the X-ray
luminosity of its host cluster. \S\ref{sec:fraction} gives
estimates for fractions of clusters with a central NVSS source and
average radio powers as functions of redshift and X-ray luminosity.
The evolution of cluster radio power is discussed in \S\ref{sec:RLF}.
The energy per particle deposited in groups and clusters since
redshift 2 is estimated in \S\ref{sec:jp}. \S\ref{sec:jp_discussion} contains some discussion of the calculation of average AGN jet power and \S\ref{sec:summary} is the  summary. We adopt a $\Lambda$CDM cosmology with $h_0=0.7$, $\Omega_{\Lambda}=0.7$, and $\Omega_m = 0.3$.

\section{AGN feedback: mechanical powers of jets} \label{sec:equation} 

X-ray cavities provide clear evidence of the interaction between AGN
jets and the hot atmospheres of clusters.  Power from AGN can be
distributed into the ICM through several channels \citep[reviewed
  in][]{mcnamara07,mcnamara12}, e.g., shock fronts
\citep[][]{nulsen05a,nulsen05b}, sound waves \citep[][]{fabian06}
driven by AGN jets.  The minimum energy required to create a cavity
can be estimated using simple assumptions.  The enthalpy, $H$,
of a cavity is equal to the sum of its thermal energy and the work
required to excavate it under constant pressure.  As discussed
  elsewhere \citep[e.g.,][]{mcnamara12}, the cavity enthalpy can only
  underestimate the total energy deposited by an expanding radio lobe,
  which might, for example, have experienced significant cosmic ray
  leakage or large adiabatic losses, particularly by driving strong
  shocks.  Thus, cavity enthalpies provide a lower limit on the energy
  distributed to the ICM from an AGN jet.
If the plasma
filling the cavity is predominantly relativistic, the enthalpy is $H = 4pV$,
where $p$ is the pressure and $V$ is the volume of the cavity.
Assuming a time scale, $\tau$, to inflate the cavity, the mean power
required to inflate the cavity is at least $\pcav\simeq 4pV/\tau$,
which provides an estimate of the jet power.  The time scale, $\tau$,
is commonly estimated using the terminal velocity of the buoyantly
rising bubbles \citep[e.g.,][]{birzan04, birzan08, dunn05}.  Such
measurements of the jet power require deep, high resolution X-ray data
to determine the volume and pressure of the cavities, so that the
measurements cannot currently be conducted for a large statistical
sample.  Nevertheless, a correlation between the radio power of the
AGN and the jet power was demonstrated by \citet{birzan04} and
improved by \citet{birzan08}, \citet{cavagnolo10},
and \citet{osullivan11}. Using this correlation, we can estimate the
minimum power necessary to inflate radio lobes from the radio power of
the central AGN.  This procedure provides a practical means to
estimate the energy deposited by radio AGN in a sample large enough
for a statistically meaningful analysis. 

In this work, we estimated jet powers using the $\phigh$ -- $\pcav$
scaling relation of \citet{cavagnolo10},
\begin{equation}
\log \pcav\ = 0.75 (\pm 0.14) \log \phigh + 1.91 (\pm 0.18),
\label{eqn:high} 
\end{equation}
where $\pcav$ is in units of $10^{42} \ergs$ and $\phigh$ is the radio
power at 1.4 GHz in units of $10^{40} \ergs$. The scatter in the
correlation between jet power and radio luminosity is $\shigh
=0.78$\,dex.  Although measurement errors contribute to this scatter,
it is dominated by intrinsic variations in radio source properties
\citep{birzan08}.  The relationship in Equation~(\ref{eqn:high}) is
determined over 7 decades in $\phigh$ ($10^{37}$ -- $10^{44}\ergs$),
for systems ranging from the nuclear radio sources of Brightest Cluster Galaxies (BCGs) in cooling
core clusters with $L_{\rm X}$ up to $10^{45}\ergs$ to the low-power
radio sources in galaxy groups, with $L_{\rm X}$ of
approximately $10^{43}\ergs$.  Note that there are only three sources
in the sample of \citet{cavagnolo10} with $\phigh > 10^{42}\ergs$ and
the relation in Equation~(\ref{eqn:high}) may overestimate
\pcav{} for them.  This is discussed further in
\S\ref{sec:sample_cavpow}.

In contrast to the cavity powers, ``beam'' powers of radio AGN
have been estimated based only on radio data \citep[e.g.,][and the
  references therein]{odea09, daly12, antognini12}.  Cavity powers and
``beam'' powers provide largely complementary means to estimate the
AGN jet power, since cavities are mostly associated with FR\,I radio
sources, whereas the beam powers are only measured for FR\,II radio
sources \citep[][]{FR}.  The two recent papers \citet{daly12} and
\citet{antognini12} discuss the relationship between beam power and
radio power for FR\,II sources.  
   The slopes they find for this relationship ($0.84\pm0.14$ in
  \citealt{daly12} and $0.95\pm0.03$ in \citealt{antognini12}) are
  steeper than given by Equation~(\ref{eqn:high}).  Nevertheless,
  because of the large scatter in these relations, their slopes differ
  from that of \citet{cavagnolo10} in Equation~\ref{eqn:high} by less
  than 2$\sigma$.  As discussed by \citet{antognini12} and
  \citet{daly12}, their relations are also consistent with
  Equation~(\ref{eqn:high}) for the range of radio powers where they
  overlap.  The difference reflects the relatively high radio
  efficiencies of FR\,II sources.  
  Here, we use Equation~(\ref{eqn:high}) to determine jet powers,
  since radio sources in clusters are mostly the lower powered FR\,I
  types.  Note that for the low end of their power range, $P_{1.4} =
  10^{23}\rm\ W\ Hz^{-1}$, the fit of \citet{antognini12} would give
  jet powers almost an order of magnitude lower than
  Equation~\ref{eqn:high}.  Since the enthalpy based estimates of jet
  power used to obtain Equation~\ref{eqn:high} are, if anything, low,
  our approach is only likely to underestimate jet powers.

\begin{figure}[h] 
\plotone{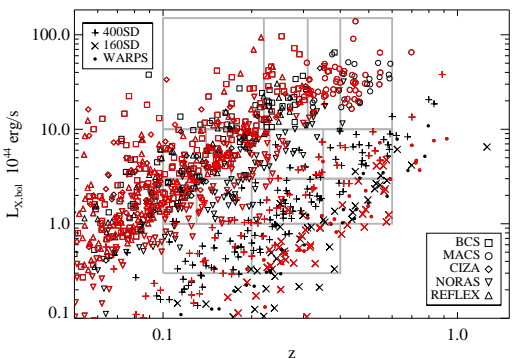}
\figcaption[]{Bolometric X-ray luminosity versus redshift for clusters
  in the eight cluster surveys used here.  Bolometric X-ray
  luminosities are derived from $0.5 - 2.0$\,keV \textit{ROSAT} PSPC
  luminosities, assuming the $L_{\rm X}$ -- $T$ relation of
  \citet{markevitch98}.  Symbols for the cluster surveys are given in
  the legend.  Red symbols denote clusters having an NVSS source
  projected within 250\,kpc of the cluster center.  The gray boxes
  mark the cells used to calculate average jet powers in
  \S\ref{sec:jp}.
\label{fig:xlum}}
\end{figure}

\begin{figure}[h] 
\plotone{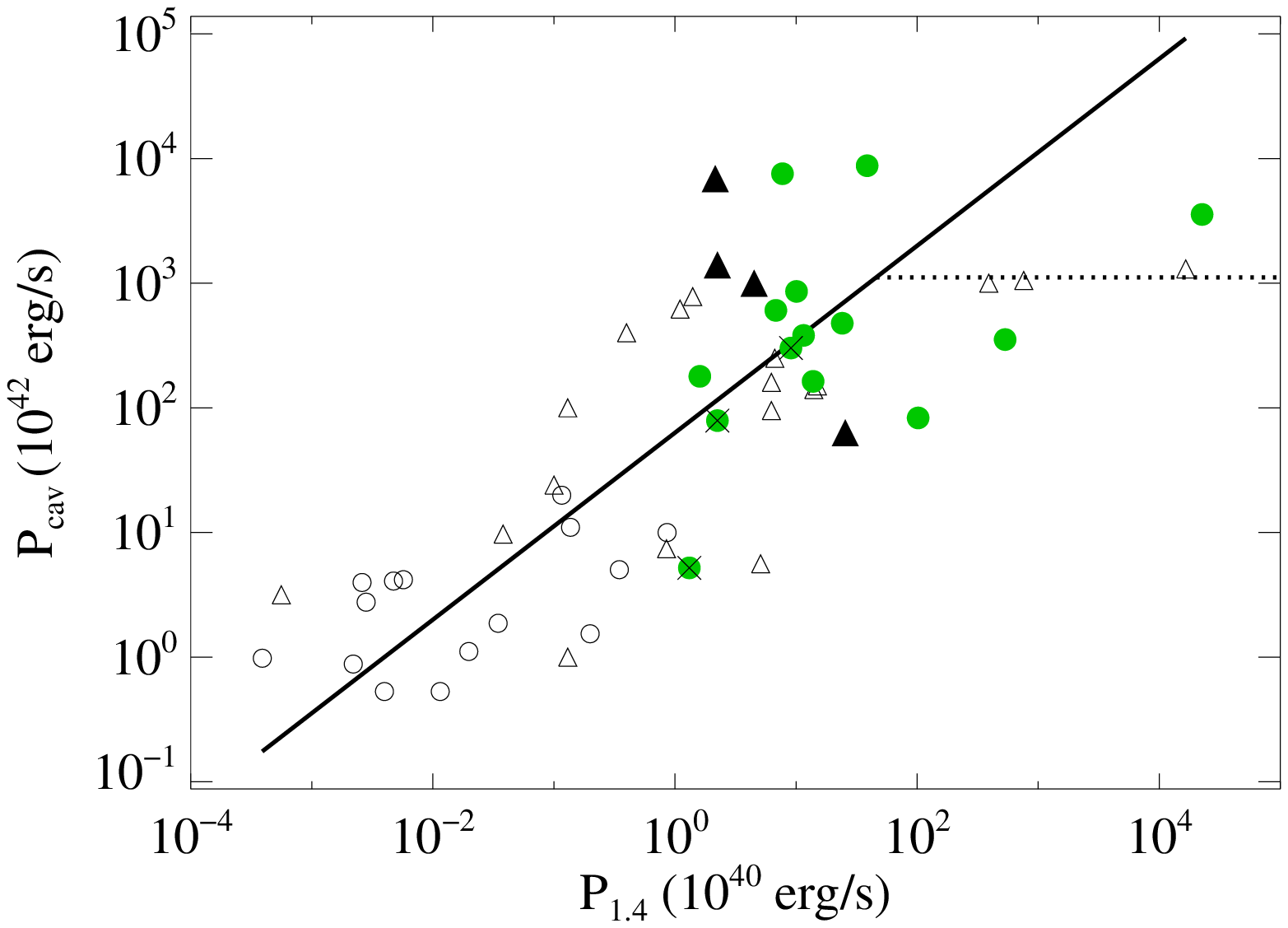}

\figcaption[]{Correlation between cavity power, $P_{\rm cav}$,
  estimated as $4pV/\tau$, and radio power, $\phigh$.  Clusters in
  our sample with known X-ray cavities are plotted as filled symbols.
  $P_{\rm cav}$ values for them are taken from \citet[filled green
    circles]{hlavacek12} and \citet[filled black triangles]{birzan08}.
  ``Possible'' cavities from the sample of \citet{hlavacek12} are
  marked with a cross to indicate that they are less secure
  detections.  Open triangles and circles denote other clusters used
  by \citet{birzan08} and \citet{cavagnolo10} respectively.  The solid line is the
  scaling relation of Equation~(\ref{eqn:high}).  The dotted line
  shows the saturation level discussed in \S\ref{sec:sample_cavpow}.
\label{fig:cavagnolo}}
\end{figure}

\begin{figure*}[t] 
\epsscale{1.00}
\plottwo{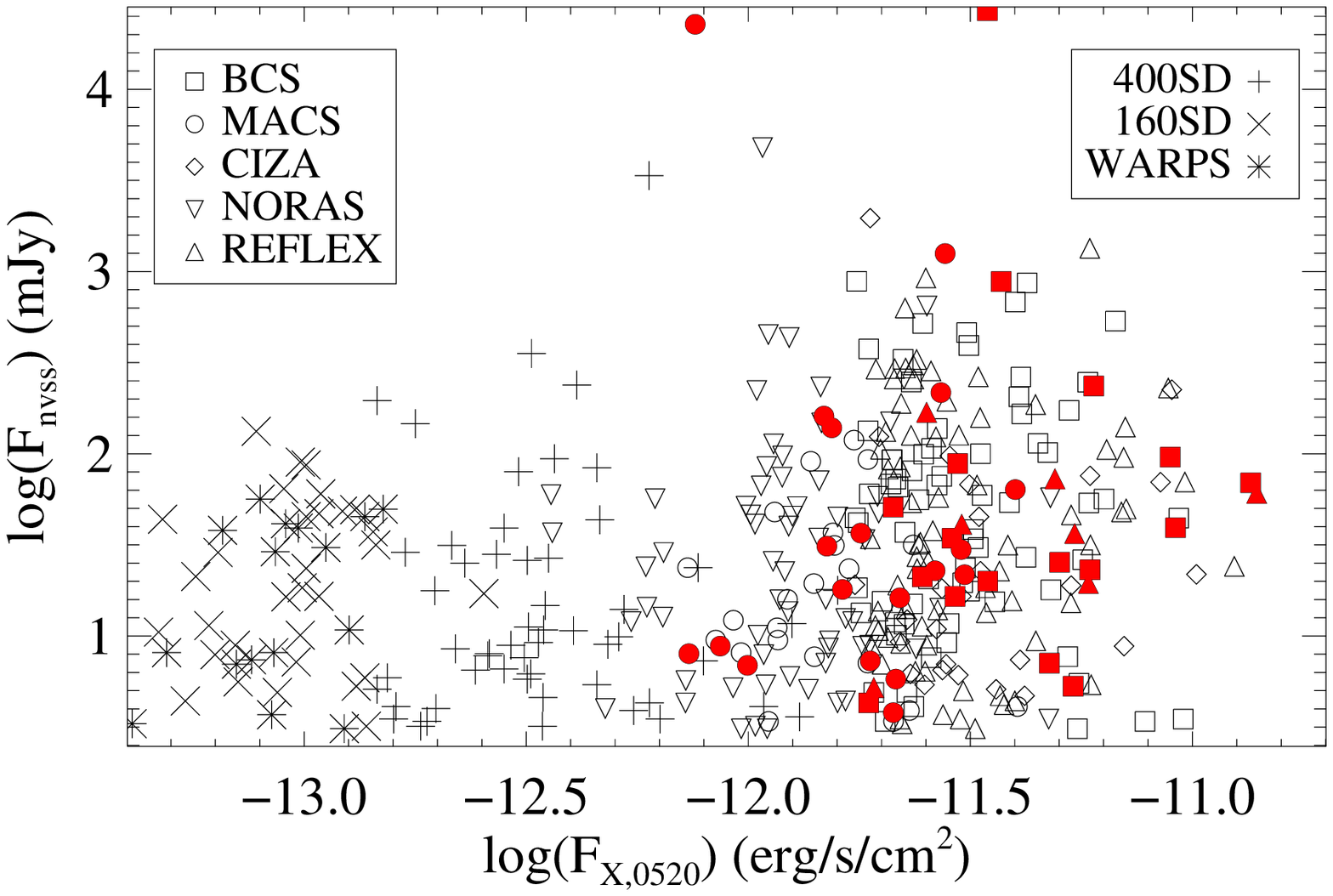}{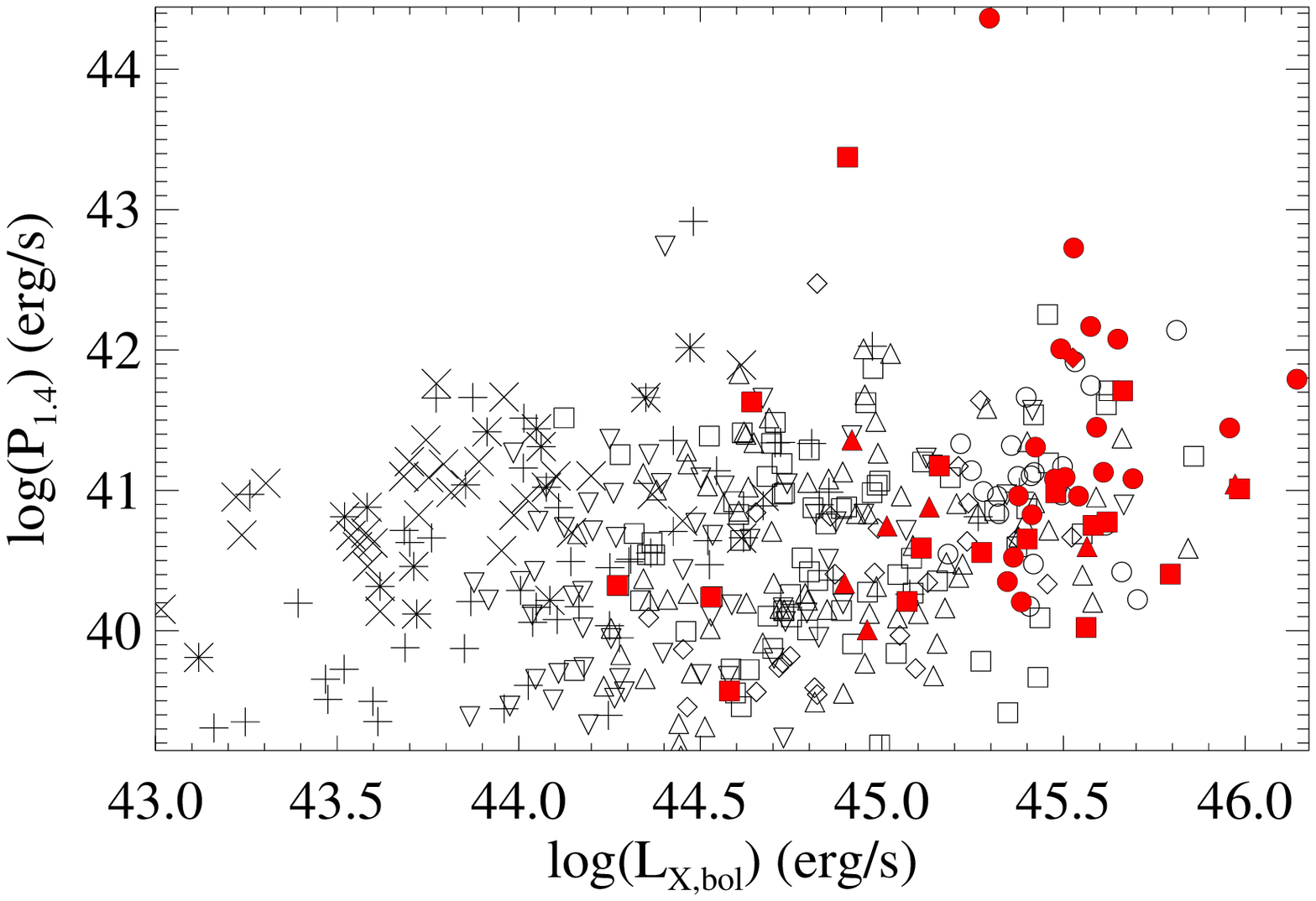}
\figcaption[]{Left: NVSS radio flux for AGN projected within 250\,kpc
  of a cluster center versus 0.5 -- 2\,keV X-ray flux of the
  cluster. Right: Radio power at 1.4 GHz for the same sources versus
  bolometric X-ray luminosity of the host.  Clusters identified as
  having strong cooling cores ($t_{\rm cool} < 1$\,Gyr) in
  \citet{cavagnolo09} are plotted in red.  Most of the remaining
  clusters are not from the sample of \citet{cavagnolo09}, so their
  cooling times are unknown.
\label{fig:correlation_flux}}
\end{figure*}

\section{The Sample}\label{sec:sample}

To extend the work of \citetalias{ma11b}, we have combined eight major
X-ray cluster surveys (Figure~\ref{fig:xlum}): the 400SD
\citep[][]{burenin07}, the 160 deg$^2$ Survey
\citep[160SD;][]{vikhlinin98a, mullis03}, the Wide Angle ROSAT Pointed
Survey \citep[WARPS;][]{warps-I,hormer08}, the MAssive Cluster Survey
\citep[MACS;][]{macs,ebeling07,ebeling10,mann12}\footnote{We only used
  the MACS subsample published in the three papers of
  \citet{ebeling07,ebeling10} and \citet{mann12}.}, the Brightest
Cluster Survey \citep[BCS;][]{bcs,ebcs}, the Clusters in the Zone of
Avoidance \citep[CIZA;][]{ebeling02,kocevski07}, the Northern ROSAT
All-Sky Cluster Survey \citep[NORAS;][]{bohringer00}, and the
ROSAT-ESO Flux Limited X-Ray Cluster Survey
\citep[REFLEX;][]{bohringer01}.

The first three surveys were compiled from serendipitously detected
clusters in targeted ROSAT PSPC observations, while the other five
cluster surveys are based on the ROSAT All Sky Survey catalogue
\citep[RASS;][]{voges99}.  Many clusters are recorded in more than one
of these surveys.  These were identified by having centroid offsets of
less than 2$\arcmin$.  Most overlapping identifications are between
the three serendipitous surveys, or NORAS and BCS.  Of the 223 160SD
clusters, 101 are included in 400SD, 40 of the 141 WARPS clusters are
included in 400SD, and 213 of the 484 NORAS clusters are included in
BCS.  Redshifts for the overlapping entries are mostly consistent
within 10\% between catalogs.  For the overlapping entries, we used
the redshift, centroid, and X-ray luminosity from the highest priority
survey.  Priority order is as listed above\footnote{Of the three
  serendipitous surveys, the most up to date, 400SD, is ranked ahead
  of 160SD, while WARPS is ranked last for its smaller sample size and
  sparser information in the public catalog \citep{warps-I,hormer08}.
  Of the all-sky surveys, the rankings of MACS, CIZA, and REFLEX are
  insignificant because they have so few overlaps with the other
  surveys.  For the two remaining surveys, BCS and NORAS, the
  redshifts and X-ray luminosities are comparably reliable.  We ranked
  BCS higher because of our familiarity with the BCS work.}, i.e. 400SD, 160SD, WARPS, MACS, BCS, CIZA, NORAS, and REFLEX.  For the
overlapping entries, the multiple redshift measurements of 18
clusters are inconsistent, with $|{\Delta z}/{z}| > 10\%$.  These
clusters were excluded from our sample to ensure the accuracy of our
redshifts and cluster identifications, although some measurement
inconsistencies seem to have been resolved in the literature
\citep[see the summary in][]{piffaretti11}.
 
To combine the X-ray luminosity measurements from different surveys,
we recalculated bolometric X-ray luminosities using the $L_{\rm X}-T$
relation from \citet{markevitch98}.  The X-ray luminosities estimated
from different entries for the same cluster were compared to examine
the consistency of flux measurements in different surveys.  The
average difference in luminosity is about 10\%.  X-ray luminosities
are used to estimate the number of particles in the clusters in
\S\ref{sec:jp} and so to derive the average AGN energy injected per
particle.  Because uncertainties in \pcav{} due to the scatter in the
scaling relation of Equation~(\ref{eqn:high}) dominate in estimates of the
average radio jet power, uncertainties in the X-ray luminosities are
not an issue in this study.

Collectively, these eight X-ray cluster surveys contain 1032 clusters
in the declination range $-38\degr <\delta < 68\degr$, an area that is
well covered by the NVSS.  The background source
density\footnote{Background source density, $\rho_{\rm bkg}$, is
  measured in an annulus extending from 2 to 5 arcmin from each
  cluster, with a flux limit of 3mJy.} for the sample is $\rho_{\rm
  bkg} > 35\,{\rm deg^{-2}}$.  We focused on clusters in the redshift
range 0.1 to 0.6.  The upper redshift limit is set by limited sampling
and the lower limit is set to avoid the complexity of radio flux
measurements for resolved sources.  After the redshift cuts, 685
clusters remain in the range of bolometric X-ray luminosities $3\times
10^{43}$ to $15\times 10^{45}$\ergs.

\subsection{Radio Sources in Clusters}\label{sec:sample_radsrc} 

Following the analysis in \citetalias{ma11b}, we cross-matched the
coordinates of the clusters with radio sources in the NVSS catalogue.
For our sample of 685 clusters, 357 have NVSS radio sources above a
flux limit of 3\,mJy projected within 250\,kpc. \citetalias{ma11b}
\citep[see also][]{lin07,best07a} showed that the density of radio
sources at the center of the clusters is much higher ($\sim
2$\,Mpc$^{-2}$) than at larger radii ($0.3$\,Mpc$^{-2}$), so the
probability that these central sources are not associated with the
clusters is small. The total expected number of background
contaminated clusters is 25, i.e., $7\%$ of the 357 clusters.  There
is little to be gained from using a smaller aperture due to the large
uncertainties in the cluster coordinates determined from ROSAT data
(see \citetalias{ma11b} for a more detailed discussion based on 400SD
clusters).

In Table~\ref{table:ratio}, we show the number of clusters in each
survey and the fraction of clusters with radio sources for two
redshift ranges, $0.1<z<0.3$, and $0.3<z<0.6$.  Here, the cluster
samples are defined by the same criteria, i.e., declination range and
NVSS background density, as the composite sample (see
\S\ref{sec:sample}), but redundant entries for a cluster in the
different surveys are not excluded.  For the lower redshift range
(column 3), the radio source fractions for the three serendipitous
surveys, at $\simeq 32\%$, are consistently lower than those for the
five all-sky surveys, which all exceed $55\%$.  For the higher
redshift range (column 6), the situation is similar, although the
variations between the all-sky surveys are greater due to large
statistical errors.  These differences between the serendipitous and
the all-sky surveys are probably due to the correlation between radio
source fraction and the X-ray luminosity of a host cluster, since the
clusters in the all-sky surveys are generally more luminous than those
in the serendipitous surveys at a given redshift. This correlation is
discussed in \S\ref{sec:fraction}.  In order to provide a fair
comparison of radio source fractions at different redshifts, f$_{\rm
  R, hif}$ in column (4) is the fraction of clusters with a central
radio source more powerful than $P_{1.4} = 3.8\times 10^{40}$\ergs,
the power cut defined for the high redshift sample.  Column (6) gives
the same fraction for the higher redshift sample, showing that these
fractions are marginally greater for the higher redshift range.

\begin{deluxetable}{lccccc}
\tablewidth{0pc}
\tablecolumns{6} 
\tablecaption{Fraction of clusters with NVSS sources\tablenotemark{a}\label{table:ratio}}
 \tablehead{
\colhead{Survey} & \multicolumn{3}{c}{$0.1<\rm z<0.3$} & \multicolumn{2}{c}{$0.3<\rm z<0.6$} \\ 
\colhead{} &  \colhead{N$_{\rm cl}$}&   \colhead{f$_{\rm R}$} &   \colhead{f$_{\rm R, hif }$ \tablenotemark{b}} & \colhead{N$_{\rm cl}$}& \colhead{f$_{\rm R, hif}$ \tablenotemark{b}} \\
\colhead{(1)} & \colhead{(2)} & \colhead{(3)} & \colhead{(4)} & \colhead{(5)} & \colhead{(6)} 
}
\startdata
   400SD    &   97        & $0.32  \pm0.06$ & $0.09\pm0.03$ &   53    &   $0.25  \pm0.1$   \\
  160SD   &   80         & $0.32  \pm0.07$ & $0.21\pm0.06$ &   63    &   $0.29   \pm0.1$   \\
  WARPS &   55         & $0.33  \pm0.09$ & $0.16\pm0.06$ &   47    &   $0.32   \pm0.1$   \\
    BCS      &  131       & $0.55  \pm0.08$ & $0.28\pm0.05$ &    9     &   $0.44   \pm0.3$    \\
   MACS   &   0            &\nodata                 & \nodata              &   65    &   $0.48   \pm0.1$   \\
  CIZA     &   35          & $0.63  \pm0.17$ & $0.29\pm0.10$ &    2     &   $0$                          \\
  NORAS &  191        & $0.55  \pm0.07$ & $0.26\pm0.04$ &  23     &   $0.13   \pm0.1$    \\
 REFLEX &  122        & $0.59  \pm0.09$ & $0.31\pm0.06$ &    7     &   $0.86   \pm0.5$    

\enddata
\tablenotetext{a}{Uncertainties in the fractions are calculated
  assuming Poisson errors in the counts.}  \\
\tablenotetext{b}{f$_{\rm R, hif}$ is the fraction of clusters having
  a central radio source with $P_{\rm1.4, lim} >
  3.8\times10^{40}$\ergs, the radio power of a 2\,mJy source at $z=0.6$.} 
\end{deluxetable}

\subsection{Cavity Powers} \label{sec:sample_cavpow} 

Some of our sample clusters were shown to have X-ray cavities by
\citet{hlavacek12} and \citet{birzan08}.  Figure~\ref{fig:cavagnolo}
shows estimates of jet power, $P_{\rm cav} = 4pV/\tau_{\rm buoy}$, for
these, plotted against their radio powers, together with the scaling
relation of Equation~(\ref{eqn:high}), and the remaining data of
\citet{birzan08} and \citet{cavagnolo10} used to establish the scaling
relation.  The new data (filled green circles) are consistent with the
scaling relation and their scatter is similar to that for the data
used to establish the scaling relation.

The scaling relation (solid line) may be overestimating \pcav{} for
the five powerful radio sources with $\phigh \geq 10^{42}$\ergs in
Figure~\ref{fig:cavagnolo} \citep[but see][and \citealt{antognini12}]{daly12}.  A similar
result is seen in Figure~1 of \citet{cavagnolo10}.  This departure could arise if a significant 
fraction of their radio synchrotron power is generated by ``hot spots'' 
which are absent from the FR I radio sources that form most of the
scaling relation \citep[e.g., 3C295;][Cygnus A;
  \citealp{wilson06}]{harris00}.  This synchrotron flux should be excluded
  before applying the scaling relation.  In some cases
\citep[e.g.,][]{forman05, lal10} shock fronts may also contribute
significantly to the power deposited into hot atmospheres.  In
principle, the energy associated with shock fronts should also be
taken into account when the \pcav{} is measured.  However,
high-quality X-ray data and careful data analysis are necessary to
detect shock fronts and to estimate the associated energy, which would
be a major undertaking for a large data set.  For consistency, values of $P_{\rm
  cav}$ in Figure~\ref{fig:cavagnolo} were estimated using only the
enthalpy and the cavity inflation time.

The number of powerful sources in Figure~\ref{fig:cavagnolo} is too small to place a strong constraint on the slope of the scaling relation at the high end.  So it is unclear whether and
to what degree we may be overestimating the jet power in these sources.  However, this issue is crucial for to making reliable estimates of \pcav{} in \S\ref{sec:jp}.  If the scaling holds,
the mean power output of short-lived but  powerful radio sources could rival 
the level of normal radio-AGN feedback over time.   
We therefore take two approaches to calculating the mean power. 
First, we excluded the most powerful radio sources, so that
the resulting average cavity power places a lower limit on the true
average cavity power.  Second, we set \pcav{} for the powerful
sources, assuming that the jet power saturates at the constant value
of $\phigh = 2.4 \times 10^{41}\,\ergs$ for high radio powers, i.e,
$\pcav (\phigh > 2.4\times 10^{41}\,\ergs) = 1.07 \times 10^{45}$\ergs
(dashed line in Figure~\ref{fig:cavagnolo}).  The saturation level is
set to the mean value of $\log\,P_{\rm cav}$ for the 5 most powerful
radio sources in Figure~\ref{fig:cavagnolo}.  The two approaches are
compared in \S\ref{sec:jp}.

\subsection{Cooling Times}\label{sec:sample:cooling} 

Central cooling times for 110 clusters in our sample were estimated
using archival \textit{Chandra} data for the ``Archive of
\textit{Chandra} Cluster Entropy Profile Tables" project
\citep[ACCEPT;][]{cavagnolo09}.  Briefly, \citet{cavagnolo09} fit
annular spectra for each cluster to determine the cooling time as a
function of the radius, assuming a profile for the cooling time of the
form
\begin{equation}
\tcool(r) = t_{c0} + t_{100} \left(\frac{r}{100 \kpc}\right)^{\alpha}, 
\label{eqn:tc0}
\end{equation}
where $t_{c0}$ and $t_{100}$ are constants.  The value for the central
cooling time, $t_{c0}$, is the cooling time \tcool{} used in this
paper.

\section{Correlation Between Radio Power and X-ray
  Luminosity} \label{sec:fluxcorr} 

The dependence of the power of a radio source on the X-ray luminosity
of its hosting cluster is shown in Figure~\ref{fig:correlation_flux}.
As found by \citetalias{ma11b}, the correlation is weak.  The Kendall
correlation coefficient for the fluxes is small, at $r = 0.16$,
although it differs from zero at high significance, the probability of
getting a value this large by chance being only $1.1 \times 10^{-4}$.
The slope of the relationship between the fluxes is $d \log F_{\rm
  NVSS} / d \log F_{\rm X, 0520} = 0.28 \pm 0.05$ and between the
powers it is $d\log\phigh / d\log L_{\rm X, bol} = 0.33 \pm 0.05$.
The surprising agreement between these slopes requires a weak
correlation between distance and luminosity for the sample
(Figure~\ref{fig:xlum}).  Since cluster X-ray luminosity increases
with mass, it follows that the radio power of a central AGN in
clusters is weakly dependent on the cluster mass.

Confining attention to clusters with cooling times shorter than
1\,Gyr, plotted in red in Figure~\ref{fig:correlation_flux}, the
Kendall correlation coefficient for the fluxes is $r = 0.3$, with a
probability of $0.04$, still significant at the 95\% level.  For these
clusters, $d \log F_{\rm NVSS} / d \log F_{\rm X, 0520} = 0.47 \pm
0.50$ and $d\log\phigh / d\log L_{\rm X, bol} = 0.51 \pm 0.30$.  While
there is marginal evidence that the radio power of a central AGN is
more strongly dependent on the X-ray luminosity in these clusters, the
correlation is weaker than expected from the work of
\citet{rafferty06}, who found a relationship between jet power and
cooling rate in clusters.  Several factors may be at work here.
First, \citet{rafferty06} use cavity powers determined from X-ray
data, a fairly diect measure, to estimate jet powers.  Using
Equation~(\ref{eqn:high}) to connect radio powers to jet powers injects
significant extra scatter into the relationship between cavity power
and cooling power.  Second, the dynamic range of X-ray luminosities in
Figure~\ref{fig:correlation_flux} is significantly smaller than that
of the cooling powers in \citet{rafferty06}, tending to bury any
correlation in the scatter.  Lastly, \citet{rafferty06} relate the
cavity power to the power radiated within the cooling radius.  If
feedback is at work, the jet power should only depend on cooling in
this region.  Although X-ray emission from within the cooling radius
can be an appreciable fraction of the total X-ray luminosity of a
cluster, the correlation is diluted by X-ray emission from larger
radii.

\subsection{Distribution of Radio Powers for Cluster Central
  AGN} \label{sec:fluxcorr_distr} 

If AGN feedback prevents cooling and star formation in cluster central
galaxies, then a high cooling rate implies a high AGN power; thus, the
radio powers of strong cooling core clusters are expected to be greater.   

\begin{figure}[h] 
\plotone{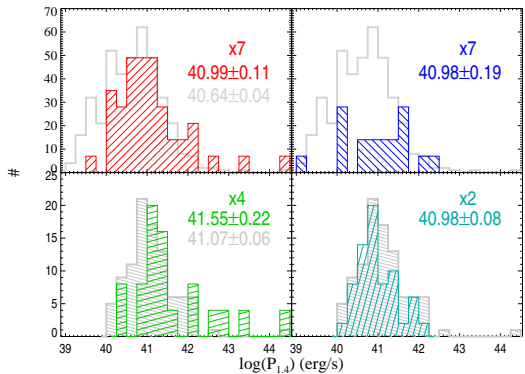}
\figcaption[]{Distributions of radio power.  The panels show
  histograms of $\phigh$ for various subsamples in different colors:
  red in the upper left panel for the cooling core clusters of
  Figure~\ref{fig:correlation_flux}, blue in the upper right panel for
  non-cooling core clusters, i.e., those with cooling times greater
  than 3\,Gyr, cyan in the lower right panel for 400SD and 160SD
  clusters at $z>0.3$, and green in the lower left panel for the clusters
  with cavities identified by \citet{hlavacek12}.  The gray histogram
  in the two upper panels shows our entire sample, as described in
  \S\ref{sec:sample}, and the filled gray histogram in the lower two
  panels show clusters in our sample with $z>0.3$.  The colored
  histograms are weighted by a factor noted in each panel to make
  comparisons easier.  Biweight means and their $95\%$ confidence
  ranges are given in the same colors as the corresponding samples.
\label{fig:p14distr}}
\end{figure}

In Figure~\ref{fig:p14distr}, we compare distributions of radio power for
the central AGN to examine whether the clusters with cooling cores or
cavities do have greater radio powers.  The red
histogram in the upper left panel shows cooling core clusters, with
$\tcool<1$\,Gyr from the ACCEPT data \citep[][]{cavagnolo09}, while the blue histogram on the
upper right shows non-cooling core clusters, with $\tcool > 3$\,Gyr
from the ACCEPT data. For comparison, the distribution for the entire
cluster sample, selected as discussed in \S\ref{sec:sample}, is
plotted in gray.  The distribution of radio powers in non-cooling core
clusters is broad, including some powerful radio sources, although the
number of clusters with $\tcool>3$\,Gyr common to our sample
and the ACCEPT sample is too small to provide
a robust result.  Further complicating matters, \citet{rafferty08} and \citet{cavagnolo08b} found that radio AGN and star formation activity at cluster centers  associated with cooling
flows are triggered when the central cooling time falls below a threshold 
of $\tcool \leq 0.5$\,Gyr. Our X-ray data are generally unable to detect such 
a threshold, making the distinction between cooling-flows and non-cooling flows
rather uncertain.

For a larger sample of non- or weak-cooling core
clusters, radio powers for 400SD and 160SD clusters with $z>0.3$ are
plotted in the cyan histogram of the lower right panel.  The 400SD
clusters at high redshifts were found to be dominated by non- or
weak-cooling core clusters by \citet[][]{santos10} and
\citet{samuele11}.  Clusters in the 160SD and 400SD cluster samples
should be similar because these two surveys used much the same
selection criteria.  To avoid bias due to the higher cut in radio
power at higher redshifts, the gray shaded histogram shows all the
clusters of our sample with $z>0.3$ for comparison.  The means of the
log of the radio power for these two samples are $\log \phigh =
40.98\pm0.08$ for the 400SD and 160SD clusters and $\log \phigh =
41.07\pm 0.06$ for the whole sample, which are consistent with one
another, showing no evidence of an offset between the two
distributions.

Despite this, the most powerful radio sources in our sample do tend to
be associated with cooling cores.  The fraction of cooling core
clusters with radio sources having $\phigh>10^{42}$\ergs is $13\pm5\%$
(6/45), compared to $10\pm7\%$ (2/19) for the
non-cooling core clusters of the upper right panel and $3\pm2\%$ (1/36) for
the 160SD and 400SD samples in the lower right panel. In summary,
radio sources in the cooling core clusters are generally as powerful
as those in the non-cooling cores, apart from the most powerful radio
sources. On the other hand, we have reliable estimates of the cooling rate for only a small sub-sample, and better coverage of deep and high resolution X-ray data are required for a more robust conclusion.

The green histogram in the lower left panel of
Figure~\ref{fig:p14distr} shows radio powers for clusters identified
with cavities by \citet{hlavacek12}\footnote{Clusters with ``clear''
  and ``potential'' cavities from \citet{hlavacek12} are included.
  Three of their ``potential" cavities show no detected radio
  source.}, with clusters from our sample at $z>0.3$ for comparison.
The clusters with cavities do have slightly greater radio powers (mean
$\log \phigh = 41.55\pm 0.22$) than clusters in our sample (mean $\log
\phigh = 41.07\pm0.06$).  Furthermore, the distribution of radio
powers for the clusters with cavities has a longer tail at high
powers.  The fraction of these clusters having a powerful radio source
($\phigh > 10^{42}\ergs$) is $30\pm13\%$ (6/20), compared to
$8\pm3\%$ (8/96) for our sample.

\begin{figure}[h] 
\plotone{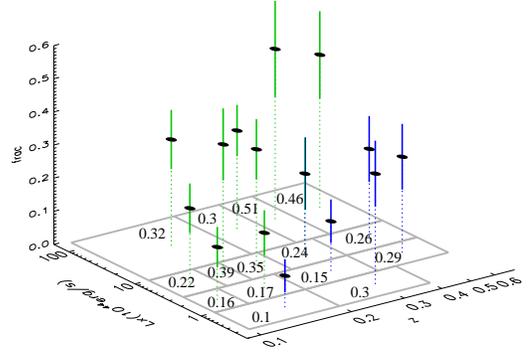}
\figcaption[]{Fraction of clusters having at least one NVSS source
  projected within 250\,kpc of the cluster center.  The NVSS sources
  are selected to have $\phigh > 3.8\times10^{40}$\ergs.  The redshift
  and X-ray luminosity bins are defined in Figure~\ref{fig:xlum}.
  Colors are used to distinguish bins dominated by clusters from the
  all-sky surveys (green) from those from the serendipitous surveys
  (blue).  Error bars show uncertainties calculated assuming Poisson
  statistics.  The fraction for each bin is noted in the bin.
\label{fig:nvssfrac}}
\end{figure}

\section{Fraction of Clusters with Radio AGN} \label{sec:fraction} 

\citetalias{ma11b} found that the probability of a more luminous and
higher-redshift 400SD cluster hosting an AGN is marginally higher than
that for a less luminous and lower-redshift 400SD cluster, at the
$1\sigma$ level.  In their relatively small sample, redshifts and
X-ray luminosities are coupled, so that the more luminous clusters
also have higher redshifts.  With our larger composite sample, we can
examine separately how the fraction of clusters matched with NVSS
radio sources depends on the X-ray luminosity and redshift, as shown
in Figure~\ref{fig:nvssfrac}.  For computing radio source fractions
here, radio sources are defined as having powers, $\phigh >
3.8\times10^{40}$\ergs, corresponding to a radio flux of 2\,mJy for a
source at $\rm z=0.6$.  The fraction of clusters having a radio source
is computed for each bin defined in Figure~\ref{fig:xlum}.  From the
figure, the fraction generally increases with both the redshift and
the X-ray luminosity of a cluster.

A significant concern is that trends in the radio fraction can be
masked by differences between the serendipitous and all-sky surveys.
For example, for the four redshift bins bounded by $z = 0.14, 0.17,
0.23, 0.35$, and 0.6, in the luminosity range $3 < L_{\rm
  X}/10^{44}\ergs <10$, the fractions of radio sources in the two
lower redshift bins, $0.14 < z < 0.23$, which are dominated by
clusters from the all-sky surveys, are higher than the fractions for
the two higher redshift bins, which are dominated by clusters from the
serendipitous surveys.  This concern is related to the persistent
question of whether the serendipitous surveys preferentially select
non-cooling core clusters, while the all-sky surveys favor cooling
core clusters \citep[cf.][]{vikhlinin06, eckert11}.  Under AGN
feedback models, cooling core clusters are more likely to host central
radio AGN \citep[e.g.,][]{rafferty06,cavagnolo08b}.  However, even if
the serendipitous and all-sky surveys differ,
Figure~\ref{fig:nvssfrac} suggests that this does not mask an
increasing trend in the radio fraction with redshift and luminosity
among the higher-redshift and more luminous clusters.  This trend can
be seen separately in the blue and green points that are dominated by
serendipitous and all-sky surveys, respectively.

\begin{figure}[h] 
\plotone{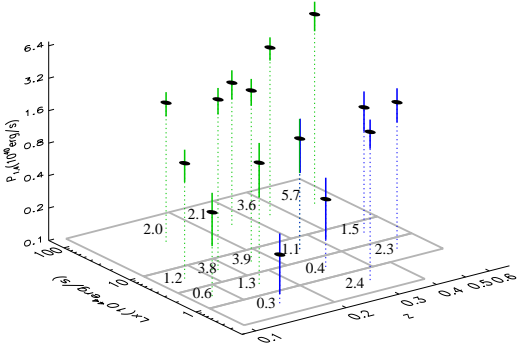}
\figcaption[]{Average radio power for sources projected within
  250\,kpc of cluster centers.  Radio sources are selected from the
  NVSS, above a radio power of $\phigh > 3.8\times10^{40}$\ergs\ for
  all redshifts.  The average radio power is recorded in each bin.
  Colors have the same meaning as in Figure~\ref{fig:nvssfrac}.
\label{fig:nvsspower}}
\end{figure}

Figure~\ref{fig:nvsspower} shows average radio powers per cluster for
the bins of Figure~\ref{fig:xlum}.  As for Figure~\ref{fig:nvssfrac},
a threshold on the radio power of $\phigh > 3.8\times10^{40}$\ergs{}
was used to avoid spurious redshift dependence in the results.  The
three most powerful radio sources in our sample, 3C~295, Hercules~A,
and 3C~288, are excluded from the averages, since they are so dominant
that they would obscure the underlying trends.  Thus, the average
radio powers in Figure~\ref{fig:nvsspower} are lower limits.

\begin{figure*}[th] 
\plottwo{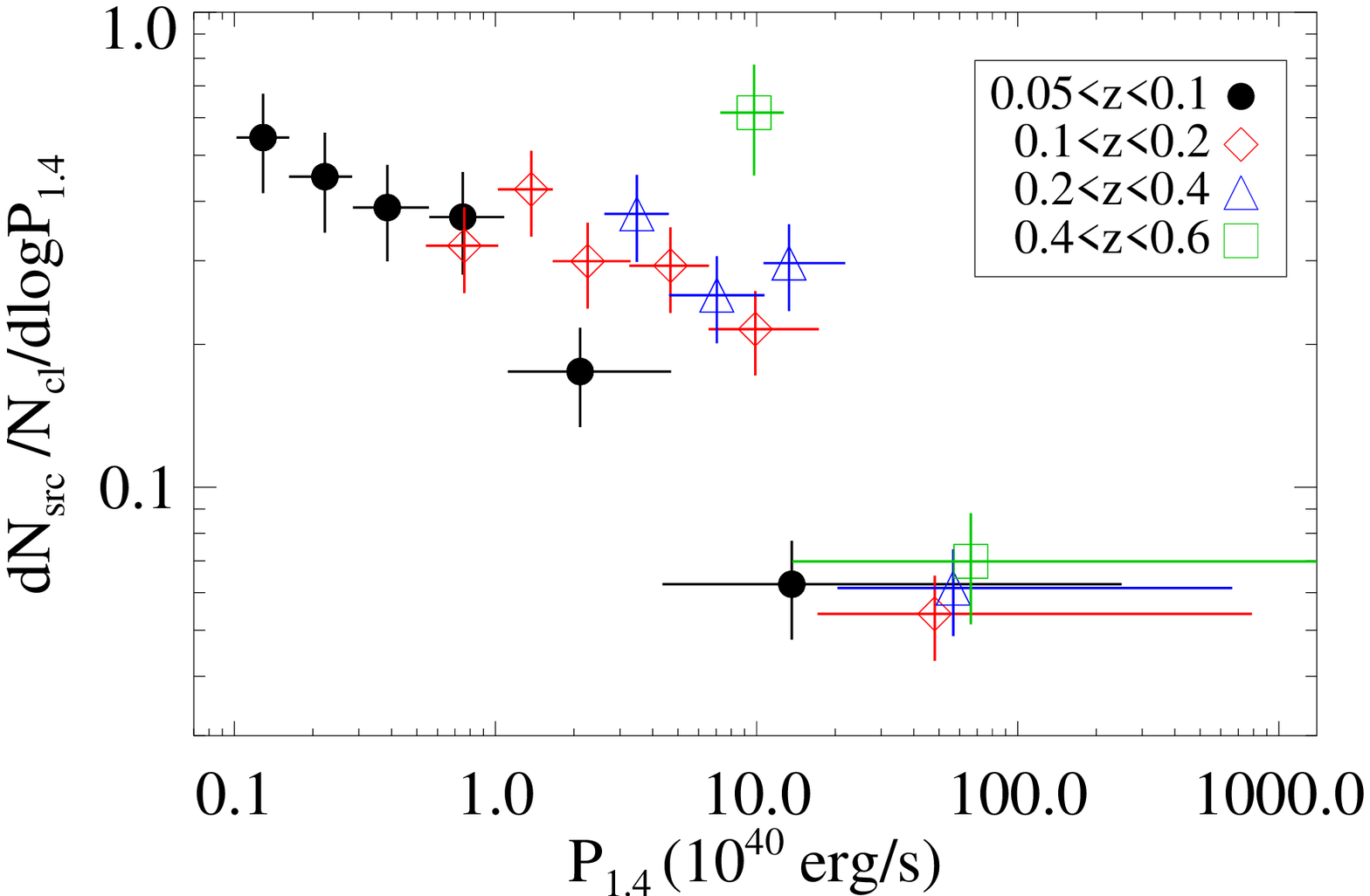}{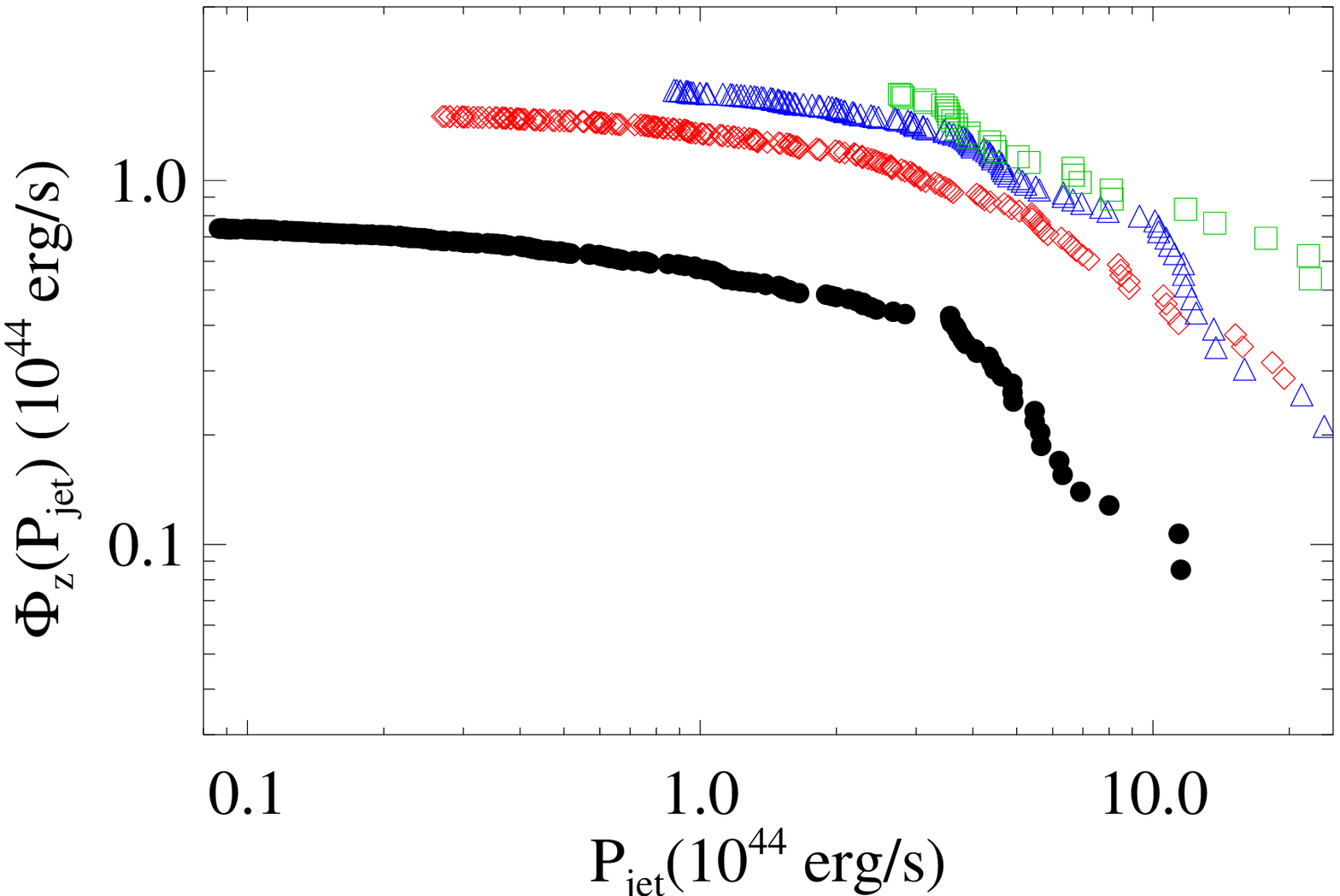}
\figcaption[]{Left:  Mean number of radio sources per cluster, per
  $\log\phigh$ as a function of NVSS radio power, $\phigh$.
  Right:  Cumulative jet power per cluster for radio jets more powerful
  than \pcav.  Details are given in \S\ref{sec:RLF}.  Both functions
  are calculated for the four redshift ranges listed in the legend of
  the left panel.  
\label{fig:RLF}}
\vspace*{1cm}
\end{figure*}

\section{Evolution of Number and Power of Cluster Central Radio
  Sources} \label{sec:RLF}  

The distribution of the number of radio sources per cluster, per
unit $\log \phigh$ is
\begin{eqnarray}
\varphi(\phigh) = {1 \over N_{\rm cl}} {dN_{\rm src}(> \phigh) \over d 
  \log \phigh}, \label{eqn:rlf0} 
\end{eqnarray}
where $N_{\rm src}(> \phigh)$ is the number of radio sources with
powers greater than $\phigh$ in the cluster population of interest and
$N_{\rm cl}$ is the number of clusters in the population.  The
expected number of background radio sources for each cluster is
subtracted from $N_{\rm src}$ and there may be more than one radio
source in a cluster.  Note that, since the normalization of $\varphi$
gives the mean number of radio sources per cluster it may be greater
than unity.  The distribution $\varphi (\phigh)$ is plotted in the
left panel of Figure~\ref{fig:RLF} for clusters in the four redshift
ranges, $0.05-0.1-0.2-0.4-0.6$.  Following \S\ref{sec:sample}, only
clusters with luminosities in the range $3\times10^{43} < L_{\rm X} <
15 \times 10^{45}$\ergs{} are included.  Values for $\varphi(\phigh)$
are shown only for powers above a threshold corresponding to the flux
limit of 3\,mJy for the redshifts $z=0.1, 0.2, 0.4$, and 0.6.  The
distribution, $\varphi(\phigh)$, increases with redshift, evolving
more significantly at the higher power end, consistent with previous
findings \citep[e.g.,][]{galametz09, hart11}.

Closely related to $\varphi(\phigh)$, we define $\phi (\pcav)$,
the number of radio jets per cluster per unit $\log \pcav$, by
replacing the radio power $\phigh$ in Equation~(\ref{eqn:rlf0}) with
\pcav\ calculated from Equation~(\ref{eqn:high}),
\begin{equation}
\phi (\pcav) = {1 \over N_{\rm cl}} {dN_{\rm src} (> \pcav)  \over d
  \log P_{\rm jet}}.
\end{equation}
The cumulative jet power per cluster from jets more powerful than
$\pcav^{\rm lim}$ is then
\begin{eqnarray}
\Phi_{z} (\pcav^{\rm lim}) = \int_{\pcav^{\rm lim}}^{\infty}
\phi(\pcav) \pcav \, d \log \pcav.
\label{eqn:rlf}
\end{eqnarray}
This is plotted as a function of $\pcav^{\rm lim}$ in four redshift
ranges in the right panel of Figure~\ref{fig:RLF}.  Here and earlier
in Equation~(\ref{eqn:rlf0}), $d \log P$ is calculated as the Voronoi
interval for the power of each radio source, although, for
$\varphi(\phigh)$, the data are binned in $\phigh$.  Values are
plotted only for $\pcav^{\rm lim}$ above a lower limit corresponding
to the radio flux limit of 3\,mJy for the different redshift ranges.

\subsection{Correcting for the Radio Flux Limit} \label{sec:RLF_weight} 

A fair comparison of the average \pcav{} per cluster for different
redshifts requires that we use the same value of $\pcav^{\rm lim}$.
For the whole sample, that would limit us to using the jet power
corresponding to the radio power cutoff for $z = 0.6$.  This is very
restrictive and it would mean discounting the power input of many less
powerful radio sources seen at lower redshifts.  Alternatively, we can
estimate the average jet power for smaller values of $\pcav^{\rm lim}$
by applying a correction factor $w$, calculated from the form of
$\Phi_{z}$ at lower redshifts, on the assumption that the shape of
$\Phi_{z}(\pcav^{\rm lim})$ does not evolve with $z$.  This
  assumption is supported by studies of the evolution of the radio
  luminosity function in, e.g., \citet{clewley04}, \citet{sadler07},
  \citet{sommer11}, and \citet{simpson12}, which find mild evolution
  of radio galaxies with $P_{1.4} < 10^{41}$\,erg/s.  It gives the 
correction factor
\begin{eqnarray}
w(z) = \frac{\Phi_{z_0} [\tilde{P}^{\rm lim}_{\rm jet} (z_0)]}
{\Phi_{z_0} [\tilde{P}^{\rm lim}_{\rm jet}(z)]}. 
\label{eqn:weight} 
\end{eqnarray}
where $\tilde{P}^{\rm lim}_{\rm jet} (z)$ is the lower limit on the
jet power for redshift $z$, which is obtained by inserting the radio
power corresponding to the flux limit of 3\,mJy at redshift $z$ into
Equation~(\ref{eqn:high}).  The cumulative jet power, $\Phi_{z_0}$, used for
reference here is that for the redshift range of $0.05 < z < 0.1$.
For example, for the redshift bin $0.4 < z <0.6$ the correction factor
is $w (z\simeq0.5) = 1.6$, calculated for $\tilde{P}^{\rm lim}_{\rm
  jet}(z) = 2.4\times 10^{44}$\ergs\ and $\tilde{P}^{\rm lim}_{\rm
  jet}(z_0) = 8.6\times10^{42}$\ergs.  The correction factor, $w(z)$,
is used in the calculation of the average jet powers below.

\begin{figure}[b] 
\epsscale{1.2}
\plotone{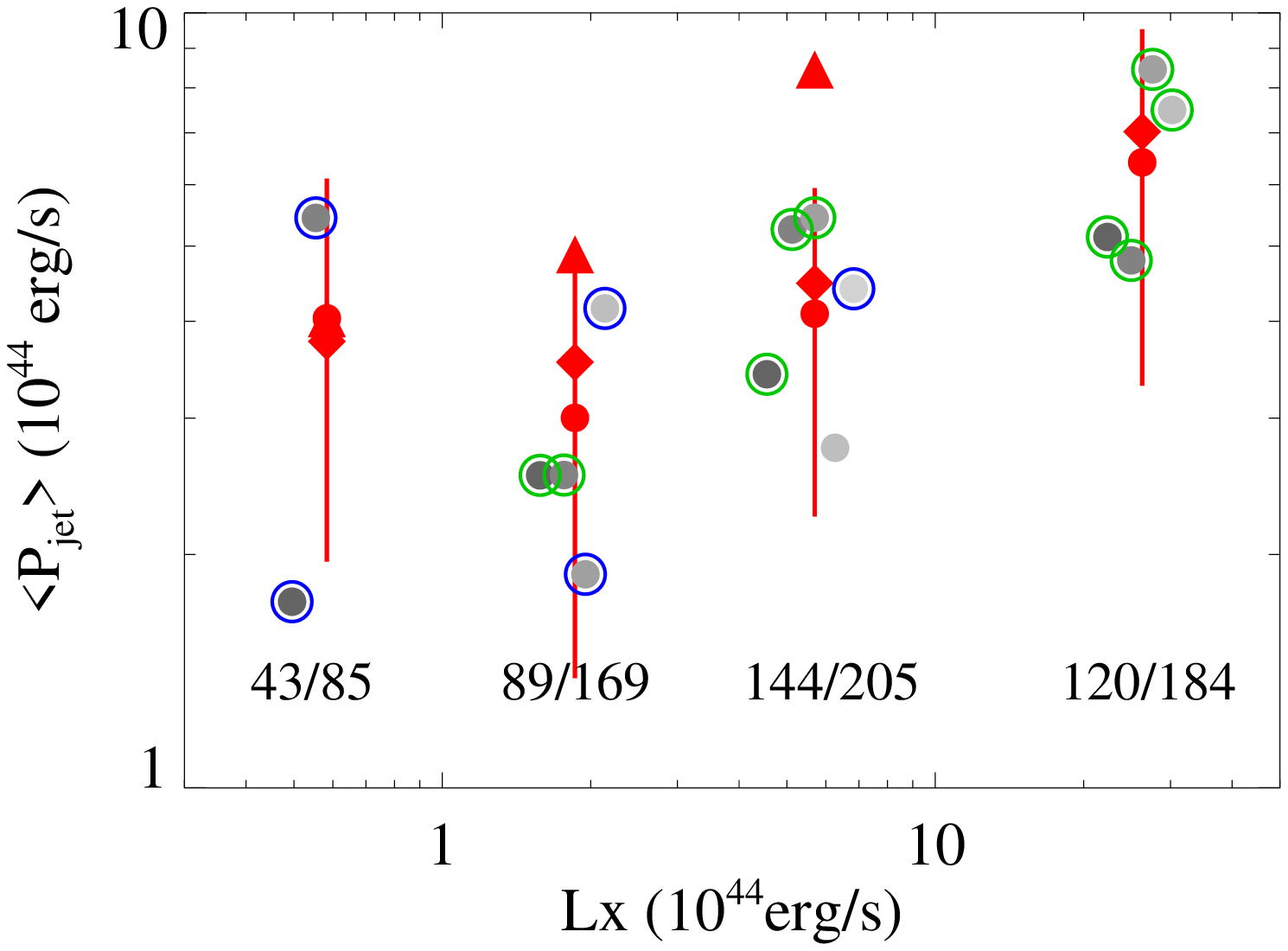}
\plotone{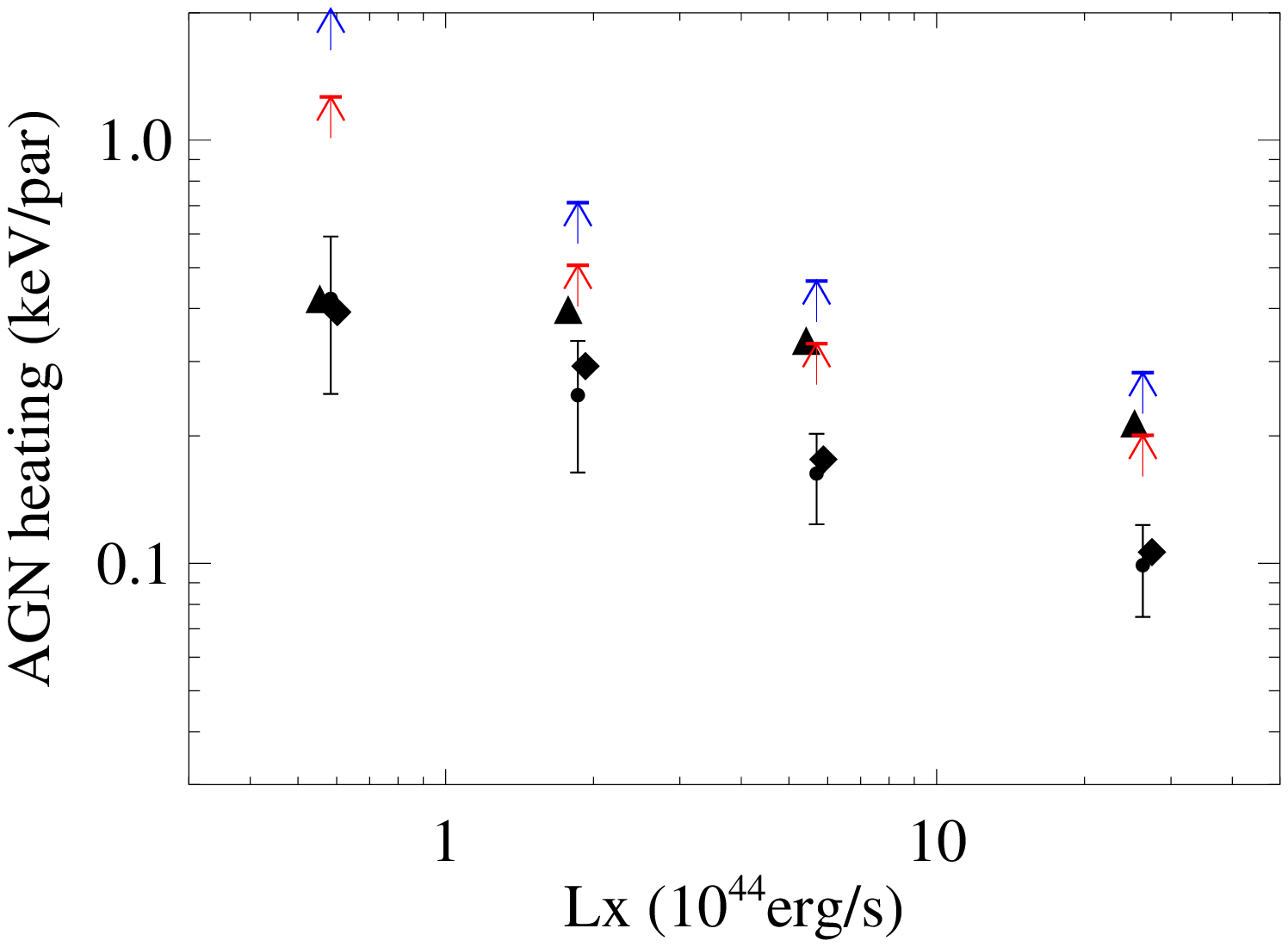}
\figcaption[]{Upper panel: Mean jet power $\langle \pcav \rangle$
  vs.\ X-ray luminosity.  Gray circles correspond to the bins in $z$
  and $L_{\rm X}$ shown in Figure~\ref{fig:xlum}.  For each range of
  $L_{\rm X}$, darker shades of gray correspond to lower redshift
  bins.  The red points with uncertainties are time averaged values,
  $\meanpcav_{\rm int}$, for the same ranges of $L_{\rm X}$ as the
  gray points.  The number of clusters hosting NVSS sources and the
  total numbers of clusters is given at the bottom of the panel for
  each range of $L_{\rm X}$.  Lower panel: Average AGN energy injected
  per particle.  The filled circles with error bars give the
  average energy per particle injected by radio jets, integrated over
  the redshift ranges marked in Figure~\ref{fig:xlum}. Jet powers are
  calculated from Equation~(\ref{eqn:high}), excluding the most powerful
  radio sources with $P_{1.4}\geq 10^{42}$\ergs. The impact of the jets is extrapolated to $z = 2$
  using two models: constant mean jet power (red arrows) and the
  linear evolution of \pcav\ discussed in \S\ref{sec:RLF} (blue
  arrows).  The effect of including \pcav\ for the most powerful radio
  sources is also shown, calculated using the saturated scaling
  relation as diamonds and the scaling relation of
  Equation~(\ref{eqn:high}) as triangles.
\label{fig:cavitypower}}
\end{figure}

\section{Average Jet Power} \label{sec:jp}

The average jet powers, \meanpcav, shown in the upper panel of
Figure~\ref{fig:cavitypower} were estimated using a Monte Carlo
method that accounts for the uncertainties in the radio fluxes and the
parameters in Equation~(\ref{eqn:high}), the distribution of radio
spectral indices, and the large intrinsic scatter (\shigh) in the
relation of Equation~(\ref{eqn:high}).  Note that the log of the arithmetic
mean of \pcav{} for a lognormal distribution is greater than the
``mean'' of its log that is given by Equation~(\ref{eqn:high}).  For each
range of $L_{\rm X}$, the redshift bins of Figure~\ref{fig:xlum} are
chosen to distribute the clusters evenly between the bins.  First,
\meanpcav\ is calculated for each bin in Figure~\ref{fig:xlum}, then
this is integrated over time for a given range of $L_{\rm X}$ to give
the time averaged mean jet power
\begin{equation}
\meanpcav_{\rm int} = \frac{\sum_{i} \left[\meanpcavzlx {w_{z_i}}
    {t_{{z}_i}} \right]}{\sum_{i} t_{z_i}},  
\label{eqn:int_pcav}
\end{equation}
where $t_{z_i}$ is the time interval for redshift bin $i$ and
$w_{z_i}$ is the correction factor calculated from
Equation~(\ref{eqn:weight}).  The full redshift range is 0.1 to 0.4 for
the lowest range of $L_{\rm X}$ and 0.1 to 0.6 for the remainder.  In
the bins for each range of $L_{\rm X}$, the evolution of \pcav\ seen
in the right panel of Figure~\ref{fig:RLF} is overwhelmed by the large
uncertainties, particularly from the scatter in the relation of
Equation~(\ref{eqn:high}).  Because of this, possible differences between
the serendipitous and all-sky surveys discussed in
\S\ref{sec:fraction} are a minor issue.

In \citetalias{ma11b}, we concluded that \meanpcav\ shows no
significant dependence on $L_{\rm X}$.  However, the limited sample
size prevented isolation of $L_{\rm X}$ from the redshift, because the
most luminous clusters in the 400SD sample are at higher redshifts.
Using the larger cluster sample here, we can break this degeneracy and
estimate the \meanpcav\ for clusters with different X-ray luminosities
over the redshift range $0.1 < z <0.6$ and the upper panel of
Figure~\ref{fig:cavitypower} shows that the mild increase of
\meanpcav\ with X-ray luminosity is not significant,
consistent with other findings \citep{gaspari11b, antognini12}.  Since
\meanpcav\ is similar for all clusters, regardless of their X-ray
luminosities, the energy input per particle from AGN is larger in less
massive clusters \citep[see also][\citetalias{ma11b}]{best07a,
    giodini10}. 

Using the $L_{\rm X}$ -- $M_{500}$ relation of \citet{vikhlinin09} and
a gas mass fraction of $0.12$, we can estimate the average gas mass within $R_{500}$, 
$\langle M_{\rm gas} \rangle$, for each luminosity range in the upper
panel of Figure~\ref{fig:cavitypower}.  Integrating the jet power over
time (cf.~Equation~\ref{eqn:int_pcav}), gives the mean energy injected
into clusters by radio AGN.  Therefore, the mean total energy per
particle injected by the radio sources is
\begin{equation}
{\rm E}_{\rm jet}=\frac{\meanpcav_{\rm int}\,t_{z, \rm int}{\mu}m_{\rm
    p}}{\langle M_{\rm gas}\rangle},  
\label{eqn:int_EperP}
\end{equation}
where $\mu = 0.59$ is the mean molecular weight, $m_{\rm p}$ is the
proton mass and $t_{z, \rm int} = \sum_i t_{z_i}$.  Here, the
integration time is limited to correspond to the redshift ranges for
the sample bins.  In principle, the energy injected by the radio
sources should be traced back to the time when BCGs formed
\citep[$z\sim2$, e.g.,][]{vandokkum01}.  Even with no AGN evolution,
extending the integration back to $z = 2$ boosts the energy injected
per particle substantially (red arrows in
Figure~\ref{fig:cavitypower}) over the values for the redshift range
of the sample (small dots).  This estimate is conservative, because
AGN are more active in the past \citep[e.g.,][]{galametz09,martini09}
and the clusters have assembled from smaller systems that may well
have contained more than one BCG.  
To allow for the evolution of the \pcav\,
  Equation~(\ref{eqn:int_EperP}) can be generalized to  
\begin{equation}
{\rm E}_{\rm jet}=\int \frac{{\mu}m_{\rm p}}{\langle M_{\rm gas}\rangle} \meanpcav_{\rm int}\, dt,
\label{eqn:int_EperP2}
\end{equation} 
where we assume that ${\langle M_{\rm gas}\rangle}$ does not depend on
the time.  The evolution of the cumulative
\pcav\ (Figure~\ref{fig:RLF} right) is modeled using a simple linear 
function, 
\begin{equation}
\meanpcav_{\rm int} \sim \Phi_{z} (\pcav) \sim {\rm A+B}t, 
\end{equation}
where the parameters (A, B) are fitted to $ \Phi_{z(t)} (\pcav)$ from
Equation~(\ref{eqn:rlf}) for $z(t) = [0.15,0.3,0.5]$, and $\pcav =
3\times10^{44}$\ergs.  This model raises the total energy input by
another factor of $50\%$ (blue arrows in 
Figure~\ref{fig:cavitypower}).  Note that our linear evolution model neglects the uncertainty in the correlation.

We discuss the interpretation of Figure~\ref{fig:cavitypower} in the
next section, \S\ref{sec:jp_discussion}. In short, the average AGN
energy input to the clusters with luminosities of $0.3 < L_{X} ({\rm
  10^{44}\,erg\,s^{-1}}) < 1.0$ can reach 1.3 -- 2 keV/particle for
ICM within $R_{500}$, depending on the details of AGN evolution.  For
the most massive clusters, with X-ray luminosities of $L_X >
10^{45}$\ergs, the average AGN input energy is also significant, at
0.2 to 0.3 keV/particle. Note that the energy input of the single AGN
outburst in the MS0735+7421 cluster is $\sim 0.25$\,keV per particle
within the central 1 Mpc \citep{gitti07b}.  It is therefore plausible
that a single, powerful AGN outburst can rival the integrated AGN
energy input over time.

\section{Discussion}\label{sec:jp_discussion} 

\subsection{AGN Energy Input}

A few points need to be addressed regarding the results in
Figure~\ref{fig:cavitypower}.  First, the average jet powers are
affected disproportionately by the most powerful radio sources, which
are the least likely to be background sources. As shown in the lower
panel of Figure~\ref{fig:cavitypower}, the average AGN energy
deposited per particle for X-ray luminous clusters would be boosted by
a factor of two assuming the scaling relation
Equation~(\ref{eqn:high}) holds for the few sources with
$\phigh>10^{42}$\ergs.  This would imply that a single, powerful radio
AGN can be as important to heating atmospheres as the integrated power
output of radio sources over time.  As discussed in section
\ref{sec:sample_cavpow}, Equation~(\ref{eqn:high}) may be
overestimating the \pcav{} for the few powerful sources with
$P_{1.4}>10^{42}$\ergs in
Figure~\ref{fig:cavagnolo}.  The saturated scaling 
relation gives average jet powers including all radio sources
(diamonds) that differ little from those obtained when the powerful
radio sources are excluded (circles).  If the scaling relation does
saturate, the small offsets between the spheres and diamonds in the
lower panel of Figure~\ref{fig:cavitypower} show that excluding the
most powerful sources makes little difference to the mean jet power
estimates.

On the other hand, our assumption that the cluster masses ($M_{500}$) remain
constant for the calculation of the mean energy injected per particle
fails to consider the hierarchical assembly of clusters.  Allowing for
cluster growth, \citet{hart11} demonstrated that jet power from
central radio AGN in clusters could increase by a factor of 10 per
particle from $z = 0.2$ back to 1.1.  At earlier times, AGN jets are
more powerful and cluster progenitors are less massive. The long term effect of preheating of these early radio AGN should be considered \citep[e.g.][]{rawlings04,shabala11}.
 Therefore,
the energy from radio jets accumulates more quickly per ICM particle
in the building blocks of present day clusters.  As less massive
clusters assemble to form more massive ones, most of the excess energy
is preserved.  By ignoring this effect, we have certainly
underestimated the total AGN energy accumulated in the ICM of massive
clusters over their histories.
One effect that might counteract this is that radio jets could break
out of the atmospheres of less massive halos \citep[e.g., the poorly
  confined sample of][]{cavagnolo10}, allowing some jet energy to
escape.  However, the huge mass of gas that will form the atmosphere
of an incipient cluster is present from the outset and very few jets
can escape from that.  Jet energy escaping from a smaller atmosphere
will be deposited in surrounding gas that is fated to collapse into
the cluster.  Unless the energy deposited by a jet is sufficient to
unbind this gas from the cluster, it inevitably collapses into the
cluster at some later time, carrying the excess energy along with it
(apart from energy lost to radiation).

Our calculations ignore energy lost from the ICM by X-ray radiation.
If we are interested in the net energy gain from radio jets, this must
be taken into account.  On the face of it, the upper panel of
Figure~\ref{fig:cavitypower} shows that the X-ray power radiated by
the clusters in our sample exceeds the average power input from the
central AGN for $L_{\rm X} \gtrsim 3 \times 10^{44}\ergs$.  Our
estimates of the mean jet power are conservative and only include
central radio sources (projected within 250 kpc), when other radio
sources may augment the total energy input significantly
\citep{shh09}.  Nevertheless, it is likely that the ICM of the most
luminous clusters suffers a net energy loss.  It should be borne in
mind that most of the X-ray power radiated by the great majority of
clusters does not originate from a cooling core.  Outside cooling
cores it will take a very long time for the energy loss to have any
noticeable impact.  Central cooling times are available in the ACCEPT
database \citep{cavagnolo09} for only 110 members of our sample,
leaving us poorly placed to examine the net effect of jets on cooling
core clusters.  Notably, Figure~\ref{fig:cavitypower} shows that, even
with our conservative estimates for the energy input from central
radio jets, clusters less luminous than $L_{\rm X} \simeq 3 \times
10^{44}\ergs$ see a net energy gain.  Thus, energy injected by central
radio AGN accumulates in lower mass clusters, so that the integrated
energy gain shown in the lower panel of the figure is mostly retained
in these systems and has a significant impact on the ICM.

As discussed in \S\ref{sec:fraction}, our sample shows increases in
the fraction of clusters with central radio AGN for increases in both
the X-ray luminosity and the redshift (Figure~\ref{fig:nvssfrac}).  It
is, therefore, surprising that we do not see a more pronounced
increase in the mean jet power with X-ray luminosity in the upper
panel of Figure~\ref{fig:cavitypower}.  The primary cause of this is
the large scatter introduced by using Equation~(\ref{eqn:high}) to
convert radio powers to jet powers.  There is good reason to believe
that mean jet powers do increase with cluster luminosities \citep[e.g.,][but see \citet{antognini12}, \citet{lin07}]{birzan04,rafferty06,sun09}.
However, the modest increase is buried by the scatter in the $\phigh -
\pcav$ relation.  It is clearly desirable to find a more accurate way
to estimate jet powers.

\subsection{Supermassive Black Hole Growth}

The integrated power output from radio-AGN at the centers of clusters
over the past $\simeq 10$ Gyr implies substantial supermassive black
hole growth.  We have estimated the accreted mass required to fuel AGN
from the integrated AGN power output over time.  We assume a
conversion efficiency between accreted mass and mechanical jet power
of $\eta=0.1$, where $P_{\rm jet}=\eta \dot Mc^2$, and we ignore
radiation loses.  We note that, although $\eta=0.1$ is commonly used in literature,  the distribution of $\eta$ is an open issue \citep[for a more detailed discussion, see][]{martinez11}. 
In some accretion models \citep[e.g.][]{benson09}, $\eta$ approaches unity at high black hole spin. However, many estimates  of $\eta$  \citep[e.g.][]{churazov05,merloni08,gaspari12b} suggest a small $\eta$, with $\eta=0.1$ near the upper limit of reasonable values for the jet efficiency. If the average efficiency is lower, the nuclear black holes need to grow even more to power these radio sources. 
Integrating the AGN mechanical energies shown in the 
bottom panel of Figure~\ref{fig:cavitypower} over $5.7$ Gyr ($z=0.6$)
gives an average accreted mass of $2-5\times10^8 M_{\odot}$ per
supermassive black hole.  Extrapolating back to $z=2.0$ over a
look-back time of about $10.5$\,Gyr, and assuming the modestly rising
AGN power discussed earlier, implies an average increase of
$6-14\times10^{8} M_{\odot}$ per supermassive black hole. Note that,
in hierarchically assembling clusters, this mass may be distributed
among several black holes. These values are comparable to the black
hole masses of BCGs inferred from black hole scaling relations
\citep[e.g.][]{lauer07}, which are thought to have been imprinted
during the quasar era.  Our result implies that normal AGN maintained
over time by hot atmospheres may be as important to supermassive black
hole growth in BCGs as earlier and, presumably, much more rapid
formation processes \citep[see the review in][]{merloni12}. It is
conceivable that normal radio-AGN activity may give rise to black hole
masses in excess of the mass expected from the $M_{\rm BH}$--$\sigma$
relation for BCGs \citep{lauer07}.  Furthermore, if $\eta$ indeed lie well
below 0.1, the inferred black hole
growth rates may be even larger, leading to the possibility of growing
ultramassive black holes in BCGs
\citep[e.g.,][]{mcnamara09,hlavacek12b}.

\section{Summary} \label{sec:summary} 

We have combined eight surveys of X-ray clusters to compile a composite
sample with 1032 clusters located in the area covered by the NVSS.
For each NVSS radio source projected within 250\,kpc of a cluster
center, we have estimated the mechanical power of its radio jet using
the scaling relation, Equation~(\ref{eqn:high}), from
\citet{cavagnolo10}.  The jet power is weakly correlated with the
X-ray luminosity of a hosting cluster, but the most powerful radio
sources, with $\phigh > 1.4\times10^{42}$\ergs, are all located in
massive, cooling core clusters.  The correlation is stronger if only
the strong cooling core clusters with $\tcool <1$\,Gyr are considered.
We have also examined the distribution of radio source powers in
cooling and non-cooling core clusters, using values of \tcool\ from
the ACCEPT project \citep[][]{cavagnolo09}.  Based on the modest
number of our sample clusters in the ACCEPT database, radio sources in
non-cooling core clusters are, in general, as powerful as those in
cooling core clusters, except that the most powerful sources mostly
appear in cooling cores.

We have examined both the average radio power of clusters and the
fraction of clusters with radio sources.  The cluster sample is large
enough to separate the dependence of the radio source fraction on
redshift and cluster X-ray luminosity and we find that it increases
moderately with both.  The average power is also larger in more
massive clusters and at higher redshifts.

Finally, we have calculated the average AGN jet power using the
scaling relation in Equation~(\ref{eqn:high}) \citep{cavagnolo10}.
This overestimates \pcav\ for the few powerful radio sources in our
sample with $\phigh\geq10^{42}$\ergs, so that the average jet power
would be dominated by these extremely powerful sources.  Two
approaches were used to solve this problem.  In the first approach,
the most powerful sources were simply excluded from the calculations,
giving a lower limit on the average jet power.  In the second
approach, \pcav\ for the powerful radio sources was determined using a
saturated version of the scaling relation, with the saturation level
set empirically and saturated jet power based on the cavity powers of
the five most powerful sources in our sample.  The average jet powers
determined using these two approaches are similar.  In the upper panel
of Figure~\ref{fig:cavitypower}, the average jet power is plotted
against the cluster X-ray luminosity.  Although the average jet power for
the most luminous clusters is higher than for less luminous
clusters, large uncertainties in their estimation make the differences
insignificant.  In general, the average jet power exceeds $3\times
10^{44}$\ergs\, even in the least luminous clusters, with $0.3< L_{\rm
  X}< 1\times10^{44}$\ergs.  Thus, the average jet power exceeds the
radiation output of the least massive sample clusters by an order of
magnitude. 

The average jet power was integrated to redshift $z=2.0$ using two
simple evolutionary models for the radio sources.  For the first
model, the radio power was taken to be constant and then the average
AGN energy injected by jets exceeds 1\,keV per particle in the least
luminous clusters, with $0.3< L_{\rm X}< 1\times10^{44}$\ergs, and
$\simeq 0.2$\,keV per particle in the most luminous clusters, with
$L_{\rm X}>10^{45}$\ergs.  Here, the number of gas particles was
calculated using the gas mass within $R_{500}$, determined from the
$L_{X}-M_{500}$ relation.  For the second model, the average jet power
was taken to be a linear function of the cosmic time and then,
integrating to $z=2.0$, the AGN energy input amounts to $\simeq 2$\,keV
per particle for clusters with $0.3< L_{\rm X}< 1\times10^{44}$\ergs
and $\simeq 0.3$\,keV per particle for clusters with $L_{\rm
  X}>10^{45}$\ergs.  

Existing X-ray data for our sample are inadequate to distinguish the
energy radiated by gas that would be significantly affected by
radiative cooling.  However, the total radiation output of the less
massive clusters is small compared to the energy input from AGN.  If
the energy injected by AGN is stored in these systems, we estimate
that the AGN energy injected since $z = 2.0$ is significant for
preheating of clusters.  If so, rather than preheating, the effect of
the AGN would be better described as ``continual heating.''  In
carrying out these calculations, we have ignored the hierarchical
assembly of clusters by assuming that the cluster masses are fixed.
Since massive clusters assembled from less massive clusters, where the
jet power per particle is larger, we expect that our estimates of the
total AGN energy accumulated in massive clusters are low.  We conclude
that continual AGN energy input in the ``radio mode'' could well
provide $>1$\,keV per particle in less massive clusters, which
approaches the excess energy required to account for observed
departures from the self-similar scaling relations that would be
expected otherwise \citep{wu00}.

Lastly, we have estimated the mass that was accreted by supermassive
black holes in BCGs to fuel their radio AGN and power their jets.
Assuming that the jet power is related to the accretion rate by
$P_{\rm jet}=0.1 \dot Mc^2$, for a typical BCG in our sample, the
nuclear black hole would have grown by about $10^{9} M_{\odot}$ since
$z = 2$.  This is comparable to black hole masses for BCGs estimated
by \citet{lauer07}, implying that that the fueling of radio AGN at the
centers of hot atmospheres may be as significant as the earlier quasar
era for the growth of supermassive black holes in BCGs.

\acknowledgements CJM and BRM are supported by Chandra Large Project Grant: G09-0140X. BRM acknowledge generous support from the Natural Sciences and Engineering Research Council of Canada.  PEJN was supported by NASA grant NAS8-03060. CJM thanks Harald Ebeling for his comments to improve the draft. This research makes use of the FIRST and NVSS radio surveys. This research has made use of the archived data and software provided by the Chandra X-ray Center (CXC) in the application packages CIAO, and Sherpa.


\begin{thebibliography}
\expandafter\ifx\csname natexlab\endcsname\relax\def\natexlab#1{#1}\fi


\bibitem[{Antognini {et~al.}(2012)Antognini, Bird, \& Martini}]{antognini12}
Antognini, J., Bird, J., \& Martini, P. 2012, ApJ, 756, 116

\bibitem[{Arnaud \& Evrard(1999)}]{arnaud99}
Arnaud, M., \& Evrard, A.~E. 1999, MNRAS, 305, 631

\bibitem[{Benson \& Babul(2009)}]{benson09}
Benson, A.~J., \& Babul, A. 2009, MNRAS, 397, 1302

\bibitem[{Best {et~al.}(2007)Best, von~der Linden, Kauffmann, Heckman, \&
  Kaiser}]{best07a}
Best, P.~N., von~der Linden, A., Kauffmann, G., Heckman, T.~M., \& Kaiser,
  C.~R. 2007, MNRAS, 379, 894

\bibitem[{B\^{i}rzan {et~al.}(2008)B\^{i}rzan, McNamara, Nulsen, Carilli, \&
  Wise}]{birzan08}
B\^{i}rzan, L., McNamara, B.~R., Nulsen, P. E.~J., Carilli, C.~L., \& Wise,
  M.~W. 2008, ApJ, 686, 859

\bibitem[{B\^{i}rzan {et~al.}(2004)B\^{i}rzan, Rafferty, McNamara, Wise, \&
  Nulsen}]{birzan04}
B\^{i}rzan, L., Rafferty, D.~A., McNamara, B.~R., Wise, M.~W., \& Nulsen, P.
  E.~J. 2004, ApJ, 607, 800

\bibitem[{B\"{o}hringer {et~al.}(2000)B\"{o}hringer, Voges, Huchra, McLean,
  Giacconi, Rosati, Burg, Mader, Schuecker, Simi\c{c}, Komossa, Reiprich,
  Retzlaff, \& Tr\"{u}mper}]{bohringer00}
B\"{o}hringer, H., {et~al.} 2000, ApJS, 129, 435

\bibitem[{B\"{o}hringer {et~al.}(2001)B\"{o}hringer, Schuecker, Guzzo, Collins,
  Voges, Schindler, Neumann, Cruddace, De~Grandi, Chincarini, Edge,
  MacGillivray, \& Shaver}]{bohringer01}
---. 2001, A\&A, 369, 826

\bibitem[{Bower {et~al.}(2006)Bower, Benson, Malbon, Helly, Frenk, Baugh, Cole,
  \& Lacey}]{bower06}
Bower, R.~G., Benson, A.~J., Malbon, R., Helly, J.~C., Frenk, C.~S., Baugh,
  C.~M., Cole, S., \& Lacey, C.~G. 2006, MNRAS, 370, 645

\bibitem[{Burenin {et~al.}(2007)Burenin, Vikhlinin, Hornstrup, Ebeling,
  Quintana, \& Mescheryakov}]{burenin07}
Burenin, R.~A., Vikhlinin, A., Hornstrup, A., Ebeling, H., Quintana, H., \&
  Mescheryakov, A. 2007, ApJS, 172, 561

\bibitem[{Cavagnolo {et~al.}(2008)Cavagnolo, Donahue, Voit, \&
  Sun}]{cavagnolo08b}
Cavagnolo, K.~W., Donahue, M., Voit, G.~M., \& Sun, M. 2008, ApJL, 683, L107

\bibitem[{Cavagnolo {et~al.}(2009)Cavagnolo, Donahue, Voit, \&
  Sun}]{cavagnolo09}
---. 2009, ApJS, 182, 12

\bibitem[{Cavagnolo {et~al.}(2010)Cavagnolo, McNamara, Nulsen, Carilli, Jones,
  \& Bîrzan}]{cavagnolo10}
Cavagnolo, K.~W., McNamara, B.~R., Nulsen, P. E.~J., Carilli, C.~L., Jones, C.,
  \& Bîrzan, L. 2010, ApJ, 720, 1066

\bibitem[{Churazov {et~al.}(2005)Churazov, Sazonov, Sunyaev, Forman, Jones, \&
  B\"ohringer}]{churazov05}
Churazov, E., Sazonov, S., Sunyaev, R., Forman, W., Jones, C., \& B\"ohringer,
  H. 2005, MNRAS, 363, L91

\bibitem[{Clewley \& Jarvis(2004)}]{clewley04}
Clewley, L., \& Jarvis, M.~J. 2004, MNRAS, 352, 909

\bibitem[{Condon {et~al.}(1998)Condon, Cotton, Greisen, Yin, Perley, Taylor, \&
  Broderick}]{nvss}
Condon, J.~J., Cotton, W.~D., Greisen, E.~W., Yin, Q.~F., Perley, R.~A.,
  Taylor, G.~B., \& Broderick, J.~J. 1998, AJ, 115, 1693

\bibitem[{Croton {et~al.}(2006)Croton, Springel, White, De~Lucia, Frenk, Gao,
  Jenkins, Kauffmann, Navarro, \& Yoshida}]{croton06}
Croton, D.~J., {et~al.} 2006, MNRAS, 365, 11

\bibitem[{Daly {et~al.}(2012)Daly, Sprinkle, O'Dea, Kharb, \& Baum}]{daly12}
Daly, R.~A., Sprinkle, T.~B., O'Dea, C.~P., Kharb, P., \& Baum, S.~A. 2012,
  MNRAS, 423, 2498

\bibitem[{Dunn \& Fabian(2006)}]{dunn06}
Dunn, R. J.~H., \& Fabian, A.~C. 2006, MNRAS, 373, 959

\bibitem[{Dunn {et~al.}(2005)Dunn, Fabian, \& Taylor}]{dunn05}
Dunn, R. J.~H., Fabian, A.~C., \& Taylor, G.~B. 2005, MNRAS, 364, 1343

\bibitem[{Ebeling {et~al.}(2007)Ebeling, Barrett, Donovan, Ma, Edge, \& van
  Speybroeck}]{ebeling07}
Ebeling, H., Barrett, E., Donovan, D., Ma, C.-J., Edge, A.~C., \& van
  Speybroeck, L. 2007, ApJ, 661, L33

\bibitem[{Ebeling {et~al.}(2000)Ebeling, Edge, Allen, Crawford, Fabian, \&
  Huchra}]{ebcs}
Ebeling, H., Edge, A.~C., Allen, S.~W., Crawford, C.~S., Fabian, A.~C., \&
  Huchra, J.~P. 2000, MNRAS, 318, 333

\bibitem[{Ebeling {et~al.}(2001)Ebeling, Edge, \& Henry}]{macs}
Ebeling, H., Edge, A.~C., \& Henry, J.~P. 2001, ApJ, 553, 668

\bibitem[{Ebeling {et~al.}(2010)Ebeling, Edge, Mantz, Barrett, Henry, Ma, \&
  van Speybroeck}]{ebeling10}
Ebeling, H., Edge, A.~C., Mantz, A., Barrett, E., Henry, J.~P., Ma, C.~J., \&
  van Speybroeck, L. 2010, MNRAS, 407, 83

\bibitem[{Ebeling {et~al.}(2002)Ebeling, Mullis, \& Tully}]{ebeling02}
Ebeling, H., Mullis, C.~R., \& Tully, R.~B. 2002, ApJ, 580, 774

\bibitem[{Ebeling {et~al.}(1996)Ebeling, Voges, B\"{o}hringer, Edge, Huchra, \&
  Briel}]{bcs}
Ebeling, H., Voges, W., B\"{o}hringer, H., Edge, A.~C., Huchra, J.~P., \&
  Briel, U.~G. 1996, MNRAS, 281, 799

\bibitem[{Eckert {et~al.}(2011)Eckert, Molendi, \& Paltani}]{eckert11}
Eckert, D., Molendi, S., \& Paltani, S. 2011, A\&A, 526, 79

\bibitem[{Evrard \& Henry(1991)}]{evrard91}
Evrard, A.~E., \& Henry, J.~P. 1991, ApJ, 383, 95

\bibitem[{Fabian(1994)}]{fabian94}
Fabian, A.~C. 1994, ARAA, 32, 277

\bibitem[{Fabian {et~al.}(2006)Fabian, Sanders, Taylor, Allen, Crawford,
  Johnstone, \& Iwasawa}]{fabian06}
Fabian, A.~C., Sanders, J.~S., Taylor, G.~B., Allen, S.~W., Crawford, C.~S.,
  Johnstone, R.~M., \& Iwasawa, K. 2006, MNRAS, 366, 417

\bibitem[{Fanaroff \& Riley(1974)}]{FR}
Fanaroff, B.~L., \& Riley, J.~M. 1974, MNRAS, 167, 31P

\bibitem[{Forman {et~al.}(2005)Forman, Nulsen, Heinz, Owen, Eilek, Vikhlinin,
  Markevitch, Kraft, Churazov, \& Jones}]{forman05}
Forman, W., {et~al.} 2005, ApJ, 635, 894

\bibitem[{Galametz {et~al.}(2009)Galametz, Stern, Eisenhardt, Brodwin, Brown,
  Dey, Gonzalez, Jannuzi, Moustakas, \& Stanford}]{galametz09}
Galametz, A., {et~al.} 2009, ApJ, 694, 1309

\bibitem[{Gaspari {et~al.}(2012)Gaspari, Brighenti, \& Temi}]{gaspari12b}
Gaspari, M., Brighenti, F., \& Temi, P. 2012, MNRAS, 3188

\bibitem[{Gaspari {et~al.}(2011)Gaspari, Melioli, Brighenti, \&
  D'Ercole}]{gaspari11b}
Gaspari, M., Melioli, C., Brighenti, F., \& D'Ercole, A. 2011, MNRAS, 411, 349

\bibitem[{Giodini {et~al.}(2010)Giodini, Smol\v{c}i\'{c}, Finoguenov,
  Boehringer, Bîrzan, Zamorani, Oklopčić, Pierini, Pratt, Schinnerer,
  Massey, Koekemoer, Salvato, Sanders, Kartaltepe, \& Thompson}]{giodini10}
Giodini, S., {et~al.} 2010, ApJ, 714, 218

\bibitem[{Gitti {et~al.}(2007)Gitti, McNamara, Nulsen, \& Wise}]{gitti07b}
Gitti, M., McNamara, B.~R., Nulsen, P. E.~J., \& Wise, M.~W. 2007, ApJ, 660,
  1118

\bibitem[{Hardcastle {et~al.}(2007)Hardcastle, Evans, \&
  Croston}]{hardcastle07}
Hardcastle, M.~J., Evans, D.~A., \& Croston, J.~H. 2007, MNRAS, 376, 1849

\bibitem[{Harris {et~al.}(2000)Harris, Nulsen, Ponman, Bautz, Cameron, David,
  Donnelly, Forman, Grego, Hardcastle, Henry, Jones, Leahy, Markevitch, Martel,
  McNamara, Mazzotta, Tucker, Virani, \& Vrtilek}]{harris00}
Harris, D.~E., {et~al.} 2000, ApJL, 530, L81

\bibitem[{Hart {et~al.}(2011)Hart, Stocke, Evrard, Ellingson, \&
  Barkhouse}]{hart11}
Hart, Q.~N., Stocke, J.~T., Evrard, A.~E., Ellingson, E.~E., \& Barkhouse,
  W.~A. 2011, ApJ, 740, 59

\bibitem[{Hlavacek-Larrondo {et~al.}(2012{\natexlab{a}})Hlavacek-Larrondo,
  Fabian, Edge, Ebeling, Sanders, Hogan, \& Taylor}]{hlavacek12}
Hlavacek-Larrondo, J., Fabian, A.~C., Edge, A.~C., Ebeling, H., Sanders, J.~S.,
  Hogan, M.~T., \& Taylor, G.~B. 2012{\natexlab{a}}, MNRAS, 421, 1360

\bibitem[{Hlavacek-Larrondo {et~al.}(2012{\natexlab{b}})Hlavacek-Larrondo,
  Fabian, Edge, \& Hogan}]{hlavacek12b}
Hlavacek-Larrondo, J., Fabian, A.~C., Edge, A.~C., \& Hogan, M.~T.
  2012{\natexlab{b}}, MNRAS, 424, 224

\bibitem[{Horner {et~al.}(2008)Horner, Perlman, Ebeling, Jones, Scharf, Wegner,
  Malkan, \& Maughan}]{hormer08}
Horner, D.~J., Perlman, E.~S., Ebeling, H., Jones, L.~R., Scharf, C.~A.,
  Wegner, G., Malkan, M., \& Maughan, B. 2008, ApJS, 176, 374

\bibitem[{Hudson {et~al.}(2010)Hudson, Mittal, Reiprich, Nulsen, Andernach, \&
  Sarazin}]{hudson10}
Hudson, D.~S., Mittal, R., Reiprich, T.~H., Nulsen, P. E.~J., Andernach, H., \&
  Sarazin, C.~L. 2010, A\&A, 513, 37

\bibitem[{Kaiser(1986)}]{kaiser86}
Kaiser, N. 1986, MNRAS, 222, 323

\bibitem[{Kaiser(1991)}]{kaiser91}
---. 1991, ApJ, 383, 104

\bibitem[{Kocevski {et~al.}(2007)Kocevski, Ebeling, Mullis, \&
  Tully}]{kocevski07}
Kocevski, D.~D., Ebeling, H., Mullis, C.~R., \& Tully, R.~B. 2007, ApJ, 662,
  224

\bibitem[{Lal {et~al.}(2010)Lal, Kraft, Forman, Hardcastle, Jones, Nulsen,
  Evans, Croston, \& Lee}]{lal10}
Lal, D.~V., {et~al.} 2010, ApJ, 722, 1735

\bibitem[{Lauer {et~al.}(2007)Lauer, Faber, Richstone, Gebhardt, Tremaine,
  Postman, Dressler, Aller, Filippenko, Green, Ho, Kormendy, Magorrian, \&
  Pinkney}]{lauer07}
Lauer, T.~R., {et~al.} 2007, ApJ, 662, 808

\bibitem[{Lin \& Mohr(2007)}]{lin07}
Lin, Y.-T., \& Mohr, J.~J. 2007, ApJS, 170, 71

\bibitem[{Ma {et~al.}(2011)Ma, McNamara, Nulsen, Schaffer, \&
  Vikhlinin}]{ma11b}
Ma, C.-J., McNamara, B.~R., Nulsen, P. E.~J., Schaffer, R., \& Vikhlinin, A.
  2011, ApJ, 740, 51

\bibitem[{Mann \& Ebeling(2012)}]{mann12}
Mann, A.~W., \& Ebeling, H. 2012, MNRAS, 420, 2120

\bibitem[{Markevitch(1998)}]{markevitch98}
Markevitch, M. 1998, ApJ, 504, 27

\bibitem[{Mart\'inez-Sansigre \& Rawlings(2011)}]{martinez11}
Mart\'inez-Sansigre, A., \& Rawlings, S. 2011, Monthly Notices of the Royal
  Astronomical Society, 414, 1937

\bibitem[{Martini {et~al.}(2009)Martini, Sivakoff, \& Mulchaey}]{martini09}
Martini, P., Sivakoff, G.~R., \& Mulchaey, J.~S. 2009, ApJ, 701, 66

\bibitem[{McNamara {et~al.}(2009)McNamara, Kazemzadeh, Rafferty, B\^{i}rzan,
  Nulsen, Kirkpatrick, \& Wise}]{mcnamara09}
McNamara, B.~R., Kazemzadeh, F., Rafferty, D.~A., B\^{i}rzan, L., Nulsen, P.
  E.~J., Kirkpatrick, C.~C., \& Wise, M.~W. 2009, ApJ, 698, 594

\bibitem[{McNamara \& Nulsen(2007)}]{mcnamara07}
McNamara, B.~R., \& Nulsen, P. E.~J. 2007, ARAA, 45, 117

\bibitem[{McNamara \& Nulsen(2012)}]{mcnamara12}
---. 2012, NJPh, 14, 5023

\bibitem[{McNamara {et~al.}(2000)McNamara, Wise, Nulsen, David, Sarazin, Bautz,
  Markevitch, Vikhlinin, Forman, Jones, \& Harris}]{mcnamara00}
McNamara, B.~R., {et~al.} 2000, ApJ, 534, L135

\bibitem[{Merloni \& Heinz(2008)}]{merloni08}
Merloni, A., \& Heinz, S. 2008, MNRAS, 388, 1011

\bibitem[{Merloni \& Heinz(2012)}]{merloni12}
---. 2012, ArXiv:1204.4265

\bibitem[{Mittal {et~al.}(2009)Mittal, Hudson, Reiprich, \& Clarke}]{mittal09}
Mittal, R., Hudson, D.~S., Reiprich, T.~H., \& Clarke, T. 2009, A\&A, 501, 835

\bibitem[{Mullis {et~al.}(2003)Mullis, McNamara, Quintana, Vikhlinin, Henry,
  Gioia, Hornstrup, Forman, \& Jones}]{mullis03}
Mullis, C.~R., {et~al.} 2003, ApJ, 594, 154

\bibitem[{Nulsen {et~al.}(2005{\natexlab{a}})Nulsen, Hambrick, McNamara,
  Rafferty, Birzan, Wise, \& David}]{nulsen05a}
Nulsen, P. E.~J., Hambrick, D.~C., McNamara, B.~R., Rafferty, D., Birzan, L.,
  Wise, M.~W., \& David, L.~P. 2005{\natexlab{a}}, ApJ, 625, L9

\bibitem[{Nulsen {et~al.}(2005{\natexlab{b}})Nulsen, McNamara, Wise, \&
  David}]{nulsen05b}
Nulsen, P. E.~J., McNamara, B.~R., Wise, M.~W., \& David, L.~P.
  2005{\natexlab{b}}, ApJ, 628, 629

\bibitem[{O'Dea {et~al.}(2009)O'Dea, Daly, Kharb, Freeman, \& Baum}]{odea09}
O'Dea, C.~P., Daly, R.~A., Kharb, P., Freeman, K.~A., \& Baum, S.~A. 2009,
  A\&A, 494, 471

\bibitem[{O'Sullivan {et~al.}(2011)O'Sullivan, Giacintucci, David, Gitti,
  Vrtilek, Raychaudhury, \& Ponman}]{osullivan11}
O'Sullivan, E., Giacintucci, S., David, L.~P., Gitti, M., Vrtilek, J.~M.,
  Raychaudhury, S., \& Ponman, T.~J. 2011, ApJ, 735, 11

\bibitem[{Peterson {et~al.}(2003)Peterson, Kahn, Paerels, Kaastra, Tamura,
  Bleeker, Ferrigno, \& Jernigan}]{peterson03}
Peterson, J.~R., Kahn, S.~M., Paerels, F. B.~S., Kaastra, J.~S., Tamura, T.,
  Bleeker, J. A.~M., Ferrigno, C., \& Jernigan, J.~G. 2003, ApJ, 590, 207

\bibitem[{Piffaretti {et~al.}(2011)Piffaretti, Arnaud, Pratt, Pointecouteau, \&
  Melin}]{piffaretti11}
Piffaretti, R., Arnaud, M., Pratt, G.~W., Pointecouteau, E., \& Melin, J.-B.
  2011, A\&A, 534, 109

\bibitem[{Rafferty {et~al.}(2008)Rafferty, McNamara, \& Nulsen}]{rafferty08}
Rafferty, D.~A., McNamara, B.~R., \& Nulsen, P. E.~J. 2008, ApJ, 687, 899

\bibitem[{Rafferty {et~al.}(2006)Rafferty, McNamara, Nulsen, \&
  Wise}]{rafferty06}
Rafferty, D.~A., McNamara, B.~R., Nulsen, P. E.~J., \& Wise, M.~W. 2006, ApJ,
  652, 216

\bibitem[{Rawlings \& Jarvis(2004)}]{rawlings04}
Rawlings, S., \& Jarvis, M.~J. 2004, MNRAS, 355, L9

\bibitem[{Sadler {et~al.}(2007)Sadler, Cannon, Mauch, Hancock, Wake, Ross,
  Croom, Drinkwater, Edge, Eisenstein, Hopkins, Johnston, Nichol, Pimbblet,
  de~Propris, Roseboom, Schneider, \& Shanks}]{sadler07}
Sadler, E.~M., {et~al.} 2007, MNRAS, 381, 211

\bibitem[{Samuele {et~al.}(2011)Samuele, McNamara, Vikhlinin, \&
  Mullis}]{samuele11}
Samuele, R., McNamara, B.~R., Vikhlinin, A., \& Mullis, C.~R. 2011, ApJ, 731,
  31

\bibitem[{Santos {et~al.}(2010)Santos, Tozzi, Rosati, \&
  B\"{o}hringer}]{santos10}
Santos, J.~S., Tozzi, P., Rosati, P., \& B\"{o}hringer, H. 2010, A\&A, 521, 64

\bibitem[{Santos {et~al.}(2012)Santos, Tozzi, Rosati, Nonino, \&
  Giovannini}]{santos12}
Santos, J.~S., Tozzi, P., Rosati, P., Nonino, M., \& Giovannini, G. 2012, A\&A,
  539, 105

\bibitem[{Scharf {et~al.}(1997)Scharf, Jones, Ebeling, Perlman, Malkan, \&
  Wegner}]{warps-I}
Scharf, C.~A., Jones, L.~R., Ebeling, H., Perlman, E., Malkan, M., \& Wegner,
  G. 1997, ApJ, 477, 79

\bibitem[{Shabala {et~al.}(2011)Shabala, Kaviraj, \& Silk}]{shabala11}
Shabala, S.~S., Kaviraj, S., \& Silk, J. 2011, MNRAS, 413, 2815

\bibitem[{Short {et~al.}(2012)Short, Thomas, \& Young}]{short12}
Short, C.~J., Thomas, P.~A., \& Young, O.~E. 2012, ArXiv e-prints:1201.1104

\bibitem[{Sijacki \& Springel(2006)}]{sijacki06}
Sijacki, D., \& Springel, V. 2006, MNRAS, 366, 397

\bibitem[{Sijacki {et~al.}(2007)Sijacki, Springel, Di~Matteo, \&
  Hernquist}]{sijacki07}
Sijacki, D., Springel, V., Di~Matteo, T., \& Hernquist, L. 2007, MNRAS, 380,
  877

\bibitem[{Simpson {et~al.}(2012)Simpson, Rawlings, Ivison, Akiyama, Almaini,
  Bradshaw, Chapman, Chuter, Croom, Dunlop, Foucaud, \& Hartley}]{simpson12}
Simpson, C., {et~al.} 2012, MNRAS, 421, 3060

\bibitem[{Sommer {et~al.}(2011)Sommer, Basu, Pacaud, Bertoldi, \&
  Andernach}]{sommer11}
Sommer, M.~W., Basu, K., Pacaud, F., Bertoldi, F., \& Andernach, H. 2011, A\&A,
  529, 124

\bibitem[{{Stocke} {et~al.}(2009){Stocke}, {Hart}, \& {Hallman}}]{shh09}
{Stocke}, J.~T., {Hart}, Q.~N., \& {Hallman}, E.~J. 2009, in American Institute
  of Physics Conference Series, Vol. 1201, American Institute of Physics
  Conference Series, ed. S.~{Heinz} \& E.~{Wilcots}, 206--209

\bibitem[{Sun(2009)}]{sun09}
Sun, M. 2009, ApJ, 704, 1586

\bibitem[{Sun {et~al.}(2007)Sun, Jones, Forman, Vikhlinin, Donahue, \&
  Voit}]{sun07}
Sun, M., Jones, C., Forman, W., Vikhlinin, A., Donahue, M., \& Voit, M. 2007,
  ApJ, 657, 197

\bibitem[{van Dokkum \& Franx(2001)}]{vandokkum01}
van Dokkum, P.~G., \& Franx, M. 2001, ApJ, 553, 90

\bibitem[{Vikhlinin {et~al.}(2006)Vikhlinin, Kravtsov, Forman, Jones,
  Markevitch, Murray, \& Van~Speybroeck}]{vikhlinin06}
Vikhlinin, A., Kravtsov, A., Forman, W., Jones, C., Markevitch, M., Murray,
  S.~S., \& Van~Speybroeck, L. 2006, ApJ, 640, 691

\bibitem[{Vikhlinin {et~al.}(1998)Vikhlinin, McNamara, Forman, Jones, Quintana,
  \& Hornstrup}]{vikhlinin98a}
Vikhlinin, A., McNamara, B.~R., Forman, W., Jones, C., Quintana, H., \&
  Hornstrup, A. 1998, ApJ, 502, 558

\bibitem[{Vikhlinin {et~al.}(2009)Vikhlinin, Burenin, Ebeling, Forman,
  Hornstrup, Jones, Kravtsov, Murray, Nagai, Quintana, \&
  Voevodkin}]{vikhlinin09}
Vikhlinin, A., {et~al.} 2009, ApJ, 692, 1033

\bibitem[{Voges {et~al.}(1999)Voges, Aschenbach, Boller, Bräuninger, Briel,
  Burkert, Dennerl, Englhauser, Gruber, Haberl, Hartner, Hasinger, Kürster,
  Pfeffermann, Pietsch, Predehl, Rosso, Schmitt, Trümper, \&
  Zimmermann}]{voges99}
Voges, W., {et~al.} 1999, A\&A, 349, 389

\bibitem[{Voit \& Donahue(2005)}]{voit05}
Voit, G.~M., \& Donahue, M. 2005, ApJ, 634, 955

\bibitem[{Wilson {et~al.}(2006)Wilson, Smith, \& Young}]{wilson06}
Wilson, A.~S., Smith, D.~A., \& Young, A.~J. 2006, ApJL, 644, L9

\bibitem[{Wu {et~al.}(2000)Wu, Fabian, \& Nulsen}]{wu00}
Wu, K. K.~S., Fabian, A.~C., \& Nulsen, P. E.~J. 2000, MNRAS, 318, 889

\bibitem[{Young {et~al.}(2011)Young, Thomas, Short, \& Pearce}]{young11}
Young, O.~E., Thomas, P.~A., Short, C.~J., \& Pearce, F. 2011, MNRAS, 413, 691


\end{thebibliography}

\end{document}